\documentclass[aps,prc,preprint,amsfonts,amsmath,superscriptaddress,floatfix,nofootinbib]{revtex4-2}
\usepackage{graphicx,amsmath,epsfig}
\newcommand{\beqy}{\begin{eqnarray}}
\newcommand{\eeqy}{\end{eqnarray}}
\newcommand{\U}{\mathcal{U}}
\newcommand{\V}{\mathcal{V}}
\newcommand{\E}{\mathfrak{E}}
\newcommand{\D}{\Delta}

\newcommand{\rb}{\pmb{r}}
\newcommand{\rp}{\pmb{r^\prime}}

\newcommand{\sg}{\sigma}

\begin{document}

\title{Superfluid fraction in the crystalline crust of a neutron star: role of BCS pairing}

\author{N. Chamel}

\affiliation{Institut d'Astronomie et d'Astrophysique, CP-226, Universit\'e Libre de Bruxelles, 
1050 Brussels, Belgium}

\begin{abstract}
The breaking of translational symmetry  in the inner crust of a neutron star leads to the 
depletion of the neutron superfluid reservoir similarly to cold atomic condensates in optical 
lattices and in supersolids.  This effect is studied in the general framework of the self-consistent 
time-dependent Hartree-Fock-Bogoliubov (HFB) theory, treating the crust as a perfect crystal. The 
superfluid fraction is derived in the Bardeen-Cooper-Schrieffer approximation for 
superfluid velocities much smaller than Landau's critical velocity within the linear-response theory. 
The different assumptions made in previous studies are clarified.  Fully three-dimensional band-structure 
calculations of superfluid neutrons in a body-centered cubic lattice are carried out.  
Although the formation of Cooper pairs is essential for the occurrence of superfluidity, the superfluid 
fraction is found to be insensitive to the pairing gap, as in uniform neutron matter.  In the 
intermediate region of the inner crust at the average baryon number density 0.03~fm$^{-3}$,  
only 8\% of the free neutrons are found to participate to the superflow.  Such very low superfluid 
fraction challenges  the classical interpretation of pulsar frequency glitches and calls for  more 
systematic calculations within the full HFB approach.  
\end{abstract}


\maketitle

\section{Introduction}

As originally discussed by Sir Anthony Leggett~\cite{leggett1970}, the breaking of translational symmetry in a superfluid leads to the existence of a normal-fluid component even at zero temperature. This implies that the superfluid velocity $\pmb{V_s}(\rb,t)$ at position $\rb$ and time $t$, defined through the gradient of the phase $\phi(\rb,t)$ of the condensate 
does not coincide with the velocity $\pmb{v}(\rb,t)$ associated with the transport of mass, i.e. with the velocity satisfying the continuity equation 
\beqy \label{eq:continuity}
\frac{\partial \rho(\rb,t)}{\partial t}+\pmb{\nabla}\cdot \bigl[\rho(\rb,t) \pmb{v}(\rb,t)\big]=0\, ,
\eeqy 
where $\rho(\rb,t)$ denotes the mass density (see Refs.~\cite{ChamelAllard2019,allard2021} for discussions in the neutron-star context). 
In an isotropic medium or in a  crystal with cubic symmetry, the spatially averaged mass current in the normal-fluid rest frame can be written as\footnote{In an anisotropic medium, the superfluid density becomes a tensor.} 
\beqy 
\pmb{\bar \rho} \equiv \frac{1}{\Omega}\int d^3\rb\, \rho(\rb,t)\pmb{v}(\rb,t)= \rho_s  \pmb{\bar V_{s}}\, . 
\eeqy 
where the superfluid density $\rho_s$ is smaller than the average mass density $\bar \rho$. The depletion of the superfluid reservoir $\rho_s/\bar \rho < 1$ 
has been recently confirmed experimentally using bosonic atomic condensates in optical 
lattices~\cite{chauveau2023,tao2023} and in a bosonic condensate in which the translational invariance has been spontaneously broken~\cite{biagioni_measurement_2024} - the realization of a so called supersolid (see, e.g., Ref.~\cite{casotti2024} and references therein).    A similar effect is expected in fermionic atomic condensates~\cite{orso2007,orso2024}.

The suppression of the neutron superfluid density in the inner crust of a neutron star was first predicted and studied in Refs.~\cite{chamel2004,CCH05a,chamel2005,CCH05b,CCH06,chamel2006}, considering both cubic\footnote{Three different cubic lattice types were considered: simple cubic, face-centered cubic, and body-centered cubic.} crystals of quasispherical clusters and more exotic nuclear pasta phases 
possibly present in the deepest region of the crust.  In this extreme astrophysical environment (see,  e.g., Ref.~\cite{blaschke2018} for a recent review about dense matter in the interior of a neutron star), the spatial modulations arise through the spontaneous formation of neutron-proton clusters.  The inner crust of a neutron star
exhibits both superfluid and elastic properties (see, e.g., Ref.~\cite{pethick2010}),  and can thus be viewed as a fermionic supersolid compound since the spontaneous breaking of translational invariance is induced by both neutrons and protons\footnote{Similarly to neutrons, protons form $^1S_0$ Cooper pairs but remain localised except possibly in the deepest layers of the crust.}.  

In the aforementioned studies, the neutron superfluid density $\rho_{n,s}$ was expressed in terms of a dynamical neutron 
effective mass defined by $m_n^\star/m_n=\rho_{n,f}/\rho_{n,s}$ with the density $\rho_{n,f}$  of free neutrons, and was estimated within the linear response 
theory implicitly considering average superfluid velocities $\bar V_{n,s}$ much smaller than Landau's velocity $V_{n,L}$ (see, e.g., Ref.~\cite{allard2023} for a recent discussion
 in the nuclear context). 
Although the formation of neutron Cooper pairs is essential for the occurrence of a superfluid phase, 
the pairing itself was expected to have a small effect on $m_n^\star$ as argued in Ref.~\cite{CCH05b} within the Bardeen-Cooper-Schrieffer (BCS) theory,
and was therefore ignored for simplicity by setting the pairing gap to zero.  
Band-structure calculations led to very large values for this dynamical effective mass, 
up to about $m_n^\star/m_n\approx 15$ in the region of quasispherical clusters at the mean baryon number density $0.03$~fm$^{-3}$~\cite{chamel2005}
implying a strong suppression of the superfluid fraction $\rho_{n, s}/\bar \rho_n\approx 6\%$ ($\bar \rho_n$ denotes the average neutron mass density) due to Bragg scattering almost independently of the crystal structure. This implies that most of the 
superfluid neutrons are entrained by the crust.  
These results were confirmed in Ref.~\cite{chamel2012} within a more realistic model (see also Ref.~\cite{chamel2017b}), and were  shown to 
have important implications for the global dynamics of neutron stars and the interpretation of pulsar frequency glitches~\cite{andersson2012,chamel2013,delsate2016} (for recent reviews about pulsar glitches, see, e.g., Refs.~\cite{antonopoulou2022,zhou2022}).

Neutron superfluidity in the crustal lattice was later studied by Martin and Urban within a purely hydrodynamical approach~\cite{martinurban2016} 
(see also Refs.~\cite{epstein1988,sedrakian1996,magierski2004} for earlier hydrodynamical studies focused on the related problem of the effective mass of 
a nuclear cluster moving in a neutron superfluid). 
Considering classical potential flows, the superfluid fraction was found to be only moderately reduced, $\rho_{n,s}/\bar \rho_n\approx 80-90\%$ at the 
same baryon density of $0.03$~fm$^{-3}$ depending on the permeability of the clusters treated as a free parameter. This approach is strictly valid in the strong coupling limit in which the 
coherence length is much smaller than the cluster size. However, this condition is generally not fulfilled (see, e.g., Ref.~\cite{Okihashi2021}), as recognized by Martin and Urban themselves
(see also the discussion in Ref.~\cite{chamel2017b}). The importance of pairing was further investigated by Watanabe and Pethick~\cite{watanabe2017} 
within the Hartree-Fock-Bogoliubov (HFB) method. The superfluid density was found to be more strongly 
suppressed than predicted by hydrodynamics, $\rho_{n,s}/\bar \rho_n\approx 60-70\%$, but much less than by band-structure calculations. 
However, many simplifying approximations were made. In particular, calculations were restricted to a linear lattice 
chain with a purely sinusoidal potential. 

In all these calculations, the crustal lattice was treated as an external periodic potential. 
Kashiwaba and Nakatsukasa~\cite{kashiwaba2019} later solved the Hartree-Fock (HF) equations self-consistently for both neutrons and protons considering 
parallel nuclear slabs %using the BCPM functional 
and obtained similar values for the dynamical effective mass (denoted in their paper by $\tilde{m}^*_z$) 
as those reported in Ref.~\cite{CCH05a}. They also introduced an alternative definition of ``free'' neutrons leading to an effective mass that is 
\emph{reduced} compared to the bare mass - a result they interpreted as ``anti-entrainment''. The ambiguity stems from the fact that neutrons are always 
free to move along the slabs. It should be emphasized, however, that this ambiguity does not arise in the crustal region of quasispherical clusters, where 
the superfluid density is expected to be the most strongly suppressed. In any case,  the superfluid density $\rho_{n,s}$ is 
well-defined independently of the specification of $\rho_{n,f}$ unlike the dynamical effective mass $m_n^\star$.  The superfluid density should thus be preferred to avoid any confusion.  
Sekizawa and collaborators~\cite{sekizawa2022} estimated the superfluid density dynamically within the self-consistent time-dependent HF theory %using the SLy4 functional 
by imposing an external force on the slabs and calculating their acceleration. They showed that this method is equivalent to static band-structure 
calculations within the numerical accuracy of their code. 
Yoshimura and Sekizawa~\cite{kenta2024} have recently solved the self-consistent time-dependent HFB equations for accelerated 
slabs in a neutron superfluid bath and found that the superfluid density is not very sensitive to the pairing gap, as was anticipated in Ref.~\cite{CCH05b}. 
Almirante and Urban have also solved the self-consistent HFB equations for slabs but considering a stationary neutron superflow and adopting a different pairing functional ~\cite{almirante2024}. The superfluid fractions obtained in this way are comparable to those of Ref.~\cite{kenta2024}. Almirante and Urban~\cite{almirante2024b} 
have recently extended their HFB calculations to a two-dimensional lattice of parallel nuclear rods. Their result in the limit of vanishing  pairing are consistent with those obtained in Ref.~\citep{CCH05a}. However, they find that pairing effects are more important in this geometry. The question 
therefore arises as to whether the strong suppression of the superfluid density found in Refs.~\cite{chamel2005,chamel2012} in the intermediate region of the inner crust 
was overestimated by the neglect of pairing. 

In this paper, fully three-dimensional calculations of the neutron superfluid fraction in the inner 
crust of a neutron star taking pairing explicitly into account in the BCS approximation are presented. The code originally used in Ref.~\cite{chamel2012} 
was completely rewritten and optimized for this purpose. The derivation of the superfluid density within the self-consistent time-dependent HFB 
framework is clarified, shedding light on implicit assumptions made in Ref.~\cite{CCH05b}. 

The microscopic model of neutron-star crust is described in Sec.~\ref{sec:micro}. After briefly reviewing the time-dependent 
HFB equations, its application to neutron star crusts is discussed in Sec.~\ref{sec:TDHFB-crust}. In Sec.~\ref{sec:linear-response}, a more rigorous 
derivation of the neutron superfluid density from the time-dependent HFB equations in the linear-response theory is presented. 
In Sec.~\ref{sec:numerical-implementation}, details of numerical calculations are provided: the numerical methods applied for computing the band structure
in Sec.~\ref{sec:band-structure}, 
as well as the integrations in the first Brillouin zone in Sec.~\ref{sec:BZ-integrations} and on the Fermi surface in Sec.~\ref{sec:FS-integrations}. 
Results are discussed in Sec.~\ref{sec:results}, first in the weak-coupling approximation in Sec.~\ref{sec:weak-coupling} and including BCS pairing in 
Sec.~\ref{sec:BCS}. The role of the vector mean-field potential is analysed in Sec.~\ref{sec:I}. Concluding remarks are given in Sec.~\ref{sec:conclusion}. 

\section{Microscopic model}
\label{sec:micro}

\subsection{Time-dependent Hartree-Fock-Bogoliubov theory}
\label{sec:TDHFB}

The time-dependent HFB theory~\cite{blaizot1986} provides a general quantum mechanical framework 
for studying the dynamics of various fermionic systems, from cold atomic gases to neutron stars
(see, e.g., Ref.~\cite{bulgac2019}). Key equations will be briefly recalled in this section. 

The energy $E$ of a nuclear matter element of volume $\Omega$ is expressed as a function of the one-body density matrix $n_q^{ij}$ and pairing tensor $\kappa_q^{ij}$ defined by 
($q=n,p$ for neutron, proton respectively)
\begin{equation}\label{eq:density-matrix-def}
	n_q^{ij}=\langle \Psi \vert c_q^{j\dagger} c_q^i \vert \Psi \rangle = n_q^{ji*} \, ,
\end{equation}
\begin{equation}
	\kappa_q^{ij}=\langle \Psi \vert  c_q^j c_q^i \vert \Psi \rangle=-\kappa_q^{ji} \, ,
\end{equation}
where $c_q^{i\dagger}$ and $c_q^i$ (using the symbols $\dagger$ and $*$ for Hermitian and complex conjugations respectively) denotes the creation and destruction operators for a nucleon of charge type $q$ in a quantum state $i$ while $\Psi$ is the many-nucleon state. Decomposing the one-body density matrix and the pairing tensor as 
\beqy\label{eq:DensityMatrix}
n_q^{ij} = \sum_k  \V^{(q)*}_{ki}\V^{(q)}_{kj}\, ,
\eeqy
\beqy
\kappa_q^{ij} =\sum_k \V^{(q)*}_{ki}\U^{(q)}_{kj} \, ,
\eeqy
the time-dependent HFB equations read~\cite{blaizot1986} ($\delta_{ij}$ is the Kronecker symbol)
\beqy
i\hbar\frac{\partial \U^{(q)}_{ki}}{\partial t}=\sum_{j}\bigl[(h_q^{ij}-\mu_q\delta_{ij})\U^{(q)}_{kj}+\D_q^{ij}\V^{(q)}_{kj}\bigr]\, ,
\label{eq:TDHFB-Uij}
\eeqy
\beqy
i\hbar\frac{\partial \V^{(q)}_{ki}}{\partial t}=\sum_{j}\bigl[-\D_q^{ij*}\U^{(q)}_{kj}-(h_q^{ij*}-\mu_q \delta_{ij})\V^{(q)}_{kj}\bigr] \, ,
\label{eq:TDHFB-Vij}
\eeqy
where $\mu_q$ denotes the chemical potential. The matrices  $h_q^{ij}$ and  $\D_q^{ij}$ of the single-particle Hamiltonian and the pair potential, respectively, are defined as 
\beqy\label{eq:Hamiltonian-matrix}
h_q^{ij}=\frac{\partial E}{\partial n_q^{ji}}=h_q^{ji*}
\, , 
\eeqy
\beqy\label{eq:pairing-matrix}
\D_q^{ij}=\frac{\partial  E}{\partial \kappa_q^{ij*}}=-\D_q^{ji}
\, .
\eeqy

For the Skyrme type functional considered here~\cite{bender03}, the energy $E$ depends on $n_q^{ij}$ and $\kappa_q^{ij}$ only through the following local densities and currents:
(i) the nucleon number density at position $\rb$ and time $t$, 
\beqy\label{eq:local-density}
n_q(\rb,t)=\sum_{\sg=\pm 1} n_q(\rb,\sg;\rb,\sg;t)
\eeqy

(ii) the kinetic-energy density (in units of $\hbar^2/2m_q)$ at position $\rb$ and time $t$, 
\beqy\label{eq:kinetic-density}
\tau_q(\rb,t)=\sum_{\sg=\pm 1}\int\text{d}^3\rp\; \delta (\rb-\rp) \pmb{\nabla}\cdot\pmb{\nabla^\prime} n_q(\rb,\sg;\rp,\sg;t)
\eeqy

(iii) the momentum density (in units of $\hbar$) at position $\rb$  and time $t$,
\beqy
\label{eq:momentum-density}
\pmb{j_q}(\rb,t)=-\frac{ i}{2}\sum_{\sigma=\pm 1}\int\,{\rm d}^3\rp\,\delta(\rb-\rp) (\pmb{\nabla} -\pmb{\nabla^\prime})n_q(\rb, \sigma; \rp, \sigma;t)\, . 
\eeqy 

(iv) the abnormal density at position $\rb$ and time $t$,
\beqy
\widetilde{n}_q(\rb,t)=\sum_{\sg = \pm 1}\widetilde{n}_q (\rb,\sg ; \rb,\sg;t )\, .
\eeqy 
In general,  Skyrme functionals depend on other densities and currents, such as the spin current vector $\pmb{J_q}(\rb,t)$ responsible for the spin-orbit coupling.  As in previous HFB calculations~\cite{kenta2024,almirante2024,almirante2024b}, this coupling is ignored.  It is very small in the inner crust of a neutron star  (see, e.g. Ref.~\cite{pearson2022}) and was shown to alter the neutron superfluid density at the third significant digit at most~\cite{chamel2005}. 

The particle and pair density matrices in coordinate space are respectively defined by~\cite{dobaczewski1984}
\begin{equation}
	n_q(\rb, \sigma; \rp, \sigma^\prime; t) = <\Psi\vert c_q(\rp,\sigma^\prime; t)^\dagger c_q(\rb,\sigma; t)\vert \Psi>\, ,
\end{equation}
\begin{equation}
	\widetilde{n}_q(\rb, \sigma; \rp, \sigma^\prime ;t) = -\sigma^\prime <\Psi\vert c_q(\rp,-\sigma^\prime; t) c_q(\rb,\sigma ; t)\vert \Psi>\, , 
\end{equation}
where $c_q(\rb,\sigma;t)^\dagger$ and $c_q(\rb,\sigma;t)$ are respectively the creation and destruction operators for a nucleon at position $\rb$ with spin $\sigma$ at time t.

Introducing the quasiparticle wavefunctions
\begin{equation}\label{eq:psi1}
\psi^{(q)}_{1k}(\rb,\sg; t)=\sum_j \varphi^{(q)}_j(\rb,\sg)\, \U^{(q)}_{kj} \, ,
\end{equation}
\begin{equation}\label{eq:psi2}
\psi^{(q)}_{2k}(\rb,\sg; t)=\sum_j (-\sg)\varphi^{(q)}_j(\rb,-\sg)^*\, \V^{(q)}_{kj} \, ,%=\sum_j \varphi_{\bar j}(\rb,\sg)\, \V_{kj} \, ,
\end{equation}
where $\varphi^{(q)}_i(\rb,\sg)$ denote single-particle basis wavefunctions with a set of quantum numbers $i$, the time-dependent HFB equations can be written in coordinate space as 
\beqy
{\rm i}\hbar\frac{\partial}{\partial t} \begin{pmatrix} \psi^{(q)}_{1k}(\rb,\sigma; t) \\ \psi^{(q)}_{2k}(\rb,\sigma; t)\end{pmatrix} =\begin{pmatrix}h_q(\rb;t)-\mu_q & \Delta_q(\rb;t) \\ \Delta_q(\rb;t)^* & - h_q(\rb;t)^* + \mu_q \end{pmatrix}\begin{pmatrix} \psi^{(q)}_{1k}(\rb,\sg; t) \\ \psi^{(q)}_{2k}(\rb,\sg; t)\end{pmatrix}\, .
\label{eq:TDHFB-Russian}
\eeqy
The single-particle Hamiltonian reads 
\begin{align}
\label{eq:Hamiltonian}
&h_q(\rb,t)= -\pmb{\nabla}\cdot\frac{\hbar^2}{2m_q^{\oplus}(\rb,t)}\pmb{\nabla}+U_q(\rb,t)-\frac{i}{2}\left[\pmb{I_q}(\rb,t)\cdot\pmb{\nabla}+\pmb{\nabla}\cdot \pmb{I_q}(\rb,t)\right]
%-{\rm i} \pmb{W_q}(\rb)\cdot\pmb{\nabla}\times\pmb{\hat \sigma}
\end{align}
with the following fields defined as
\begin{align}\label{eq:Fields1}
 &\frac{\hbar^2}{2m_q^{\oplus}(\rb,t)}=\frac{\delta E}{\delta \tau_q(\rb,t)}\,,  \qquad U_q(\rb,t)=\frac{\delta E}{\delta n_q(\rb,t)}\,, \qquad \pmb{I_q}(\rb,t)=\frac{\delta E}{\delta \pmb{j_q}(\rb,t)}\,. 
\end{align}
The pair potential is defined by
\beqy
\label{eq:pair-pot}
\D_q(\rb,t) =2\frac{\delta E}{\delta \widetilde{n}_q(\rb,t)^*}\, .
\eeqy

The effective mass $m_q^\oplus(\rb,t)$ and the vector potential $\pmb{I_q}(\rb,t)$ are intimately related as they come from the same terms in the nuclear energy density functional. Indeed, the potential energy density can only depend on the combinations $X_0(\rb,t)=n_0(\rb,t)\tau_0(\rb,t) -  j_0(\rb,t)^2$ and $X_1(\rb,t)=n_1(\rb,t)\tau_1(\rb,t) -  j_1(\rb,t)^2$, as required by Galilean invariance (see, e.g., Refs.~\cite{engel1975,dobaczewski1995}). The subscripts $0$  and $1$ refer to isoscalar and isovector quantities respectively. The former (also written without any subscript) are sums over neutrons and protons (e.g. $n_0\equiv n=n_n+n_p$) while the latter are differences between neutrons and protons (e.g. $n_1=n_n-n_p$). 
Therefore, the vector potential arises whenever there is some current and the Skyrme effective mass differs from the bare mass.

The abnormal density $\widetilde{n}_q(\rb,t)$ is the local order parameter of the superfluid phase. In general, the order parameter is complex and the gradient of its phase $\phi_q(\rb,t)$ defines the so-called superfluid velocity as follows (see, e.g., Ref.~\cite{allard2021}) 
\begin{align}
	\pmb{V_{q,s}}(\rb,t)=\frac{\hbar}{2m_q} \pmb{\nabla}\phi_q(\rb,t)\, .
\end{align}
Physically, this velocity actually represents the \emph{momentum per unit mass} carried by Cooper pairs. 
The true velocity $\pmb{v_q}$ associated with mass transport~\eqref{eq:continuity}, is given by~\cite{ChamelAllard2019,allard2021} 
\beqy\label{eq:true-velocity}
\pmb{v_q}(\rb,t) =\frac{m_q}{m_q^\oplus(\rb,t)} \frac{\hbar \pmb{j_q}(\rb,t)}{\rho_q(\rb,t)}+\frac{\pmb{I_q}(\rb,t)}{\hbar} \, . 
\eeqy

In an homogeneous medium with a single nucleon species $q$, $\pmb{V_{q,s}}$ coincides with $\pmb{v_q}$, and the (total) momentum density is simply given by $\hbar \pmb{j_q}=\rho_q \pmb{V_{q,s}}$ provided $V_{q,s}$ is lower than Landau's velocity $V_{q,L}$~\cite{allard2023}. However, these simple relations do not generally hold in superfluid mixtures as in the outer core of neutron stars~\cite{ChamelAllard2019,allard2021,allard2021universe} or in presence of inhomogeneities as in the inner crust of a neutron star - the main focus of this work. 

\subsection{Application to neutron-star crusts}
\label{sec:TDHFB-crust}

The crust is assumed to be a perfect crystal characterized by primitive translation vectors $\pmb{a_1}$, $\pmb{a_2}$, and $\pmb{a_3}$. 
In the absence of any current, both the neutron single-particle Hamiltonian and the pair potential are
invariant under lattice translations, i.e., $h_n(\rb+\pmb{\ell})=h_n(\rb)$ and similarly for $\D_n(\rb)$ where $\pmb{\ell}=n_1\pmb{a_1}+n_2\pmb{a_2}+n_3\pmb{a_3}$ with $(n_1,n_2,n_3)$ arbitrary integers.  The HFB equations are solved in a sufficiently large volume $\Omega$ (containing a macroscopic number of lattice sites) 
with Born-von K\'arm\'an periodic boundary conditions.  

Let us now consider the existence of a stationary neutron superflow with a superfluid velocity (in the crust frame) averaged over the volume $\Omega$ written as 
\beqy
\pmb{\bar V_{n,s}}\equiv\frac{1}{\Omega}\int {\rm d}^3r\, \pmb{V_{n,s}}(\rb)=\frac{\hbar \pmb{Q}}{m_n} \, .
\eeqy
The single-particle Hamiltonian $h_n(\rb)$ still has the periodicity of the lattice, but the pair potential does not and can be written as 
$\D_n(\rb)=\D_{\pmb{Q}}(\rb)e^{2{\rm i}\pmb{Q}\cdot\rb}$ where $\D_{\pmb{Q}}(\rb)$ is periodic and real. 
Substituting  $i\hbar \partial/\partial t$ by the neutron quasiparticle energies $\E$, and introducing 
\begin{equation}\label{eq:psi1tilde}
\widehat{\psi}_{1}(\rb,\sg)=\psi_{1}(\rb,\sg)e^{{\rm i} \pmb{Q}\cdot \rb}\, ,
\end{equation}
\begin{equation}\label{eq:psi2tilde}
\widehat{\psi}_{2}(\rb,\sg)=\psi_{2}(\rb,\sg) e^{-{\rm i} \pmb{Q}\cdot \rb}\, ,
\end{equation}
the time-dependent HFB equations~\eqref{eq:TDHFB-Russian} for neutrons reduce to 
\beqy
\begin{pmatrix}h_{\pmb{Q}}(\rb) -\mu_n & \D_{\pmb{Q}}(\rb) \\ \D_{\pmb{Q}}(\rb)^* & - h_{\pmb{Q}}(\rb)^* + \mu_n \end{pmatrix}\begin{pmatrix} \widehat{\psi}_{1}(\rb,\sg) \\ \widehat{\psi}_{2}(\rb,\sg)\end{pmatrix}=\E\begin{pmatrix} \widehat{\psi}_{1}(\rb,\sigma) \\ \widehat{\psi}_{2}(\rb,\sigma)\end{pmatrix} \, ,
\eeqy
where $h_{\pmb{Q}}(\rb)\equiv e^{-{\rm i} \pmb{Q}\cdot \rb} h_n(\rb) e^{{\rm i} \pmb{Q}\cdot \rb}$. Now both $h_{\pmb{Q}}(\rb)$ and $\D_{\pmb{Q}}(\rb)$ are invariant under lattice translations. Therefore, the transformations~\eqref{eq:psi1tilde} and \eqref{eq:psi2tilde} correspond to a change to the frame in which $\bar V_{n,s}=0$. Note, however, that the superfluid is generally not at rest in the sense that the true velocity $\pmb{v_n}$ does not vanish.

The quasiparticle wavefunction must obey the Floquet-Bloch theorem~\cite{kittel}: 
\begin{equation}
\widehat{\psi}_{1\pmb{k},\pmb{Q}}(\rb,\sg)=\varphi_{1\pmb{k},\pmb{Q}}(\rb,\sg)e^{{\rm i} \pmb{k}\cdot \rb}\, ,
\end{equation}
\begin{equation}
\widehat{\psi}_{2\pmb{k},\pmb{Q}}(\rb,\sg)=\varphi_{2\pmb{k},\pmb{Q}}(\rb,\sg) e^{{\rm i} \pmb{k}\cdot \rb}\, ,
\end{equation}
where $\varphi_{1\pmb{k},\pmb{Q}}(\rb,\sg)$ and $\varphi_{1\pmb{k},\pmb{Q}}(\rb,\sg)$ are periodic functions. In general, they will also depend on $\pmb{Q}$. 
Going back to the crust frame, the original quasiparticle wavefunctions read 
\begin{equation}
\psi_{1\pmb{k},\pmb{Q}}(\rb,\sg)=\varphi_{1\pmb{k},\pmb{Q}}(\rb,\sg)e^{{\rm i} (\pmb{k}+\pmb{Q})\cdot \rb}\, ,
\end{equation}
\begin{equation}
\psi_{2\pmb{k},\pmb{Q}}(\rb,\sg)=\varphi_{2\pmb{k},\pmb{Q}}(\rb,\sg) e^{{\rm i}(\pmb{k}-\pmb{Q})\cdot \rb}\, .
\end{equation}

In the BCS approximation, only neutrons having opposite spins and Bloch wave vectors $\pmb{k}+\pmb{Q}$ and $-\pmb{k}+\pmb{Q}$ respectively, 
are paired. For convenience, the following shorthand notation is introduced 
\beqy
k\equiv (\pmb{k}+\pmb{Q},\sigma)\,,\qquad\qquad \bar{k}\equiv (-\pmb{k}+\pmb{Q},-\sigma)\,  .
\label{eq:NonVanishingIndices}
\eeqy
The quasiparticle wavefunctions reduce to 
\begin{equation}\label{eq:psi1-bcs}
\psi_{1k}(\rb,\sg)=\mathcal{U}_{kk}\, \varphi_{k}(\rb,\sg)\, ,
\end{equation}
\begin{equation}\label{eq:psi2-bcs}
\psi_{2k}(\rb,\sg)=-\mathcal{V}_{k\bar k}\, \sigma\varphi_{\bar k}(\rb,-\sg)^*\, .
\end{equation}
Here $\varphi_{k}(\rb,\sg)$ is the eigenstate of $h_n(\rb)$ with single-particle energy $\epsilon_k$ and is normalized as
\beqy 
\sum_\sg \int {\rm d}^3\rb \vert \varphi_{k}(\rb,\sg)\vert^2 = 1\, .
\eeqy 
Note that $\varphi_{k}(\rb,\sg)$ is a Bloch wavefunction with wave vector $\pmb{k}+\pmb{Q}$. 
Equations~\eqref{eq:psi1-bcs} and \eqref{eq:psi2-bcs} are exact in uniform neutron matter. Quite remarkably, the HFB equations remain 
formally the same in a periodic medium as in the crust of a neutron star~\cite{allard2021} 
\beqy\label{eq:HFB-homogeneous}
\begin{pmatrix}
\xi_k & \Delta_k  \\ 
\Delta_k^* & -\xi^*_{\bar{k}}
\end{pmatrix}\begin{pmatrix} \U_{kk} \\ \V_{k\bar{k}}\end{pmatrix}=\E_k\begin{pmatrix} \U_{kk} \\ \V_{k\bar{k}}\end{pmatrix}\, ,
\eeqy
where $\xi_k\equiv \epsilon_k-\mu$. 
The matrix elements $\Delta_k\equiv\Delta_{k\bar{k}}=-\Delta_{\bar k k}$ of the pair potential are given by
\beqy 
\Delta_k = -\sum_\sg \sg \int{\rm d}^3\rb\, \varphi_{k}(\rb,\sg)^* \Delta(\rb) \varphi_{\bar k}(\rb,-\sg)^* \, .
\eeqy
The solutions are readily obtained: 
\beqy\label{eq:QuasiparticleEnergy}
\E_{k} = \frac{\xi_k-\xi_{\bar k}}{2} +\sqrt{\varepsilon_k + \vert\Delta_k\vert^2}\,,
\eeqy
\beqy\label{eq:UTerms}
\vert\U_{kk}\vert^2=\frac{1}{2}\left(1+ \frac{\varepsilon_k}{\sqrt{\varepsilon_k^2 + \vert\Delta_k\vert^2}}  \right)\,,
\eeqy
\beqy\label{eq:VTerms}
\vert\V_{k\bar{k}}\vert^2=\frac{1}{2}\left(1- \frac{\varepsilon_k}{\sqrt{\varepsilon_k^2 + \vert\Delta_k\vert^2}}  \right)\,,
\eeqy
where 
\begin{equation}
\varepsilon_k\equiv \frac{\xi_k+\xi_{\bar k}}{2} = \varepsilon_{\bar k}\, .
\end{equation}

The average neutron mass density and the average neutron mass current can be respectively expressed as~\cite{ChamelAllard2019,allard2021} 
\beqy \label{eq:average-neutron-mass-density}
\bar \rho_n = \frac{1}{\Omega}\int{\rm d}^3\rb\, \rho_n(\rb) = \frac{m_n}{\Omega}\sum_k \vert\V_{k\bar{k}}\vert^2 \, ,
\eeqy 
\beqy\label{eq:average-neutron-mass-current}
\pmb{\bar \rho_n} = \frac{1}{\Omega}\int{\rm d}^3\rb\, \pmb{\rho_n}(\rb) = \frac{m_n}{\Omega}\sum_k \vert\V_{k\bar{k}}\vert^2 \pmb{v}_k\, ,
\eeqy 
where ($\pmb{\nabla}_{\pmb{k}}$ denotes the gradient operator in $\pmb{k}$-space)
\beqy 
\pmb{v}_k\equiv \frac{1}{\hbar} \pmb{\nabla}_{\pmb{k}} \epsilon_k 
\eeqy 
represents the group velocity of the single-particle state with Bloch wave vector $\pmb{k}+\pmb{Q}$. Note that the summations over the quantum 
numbers $k$ are summations over the wave vectors $\pmb{k}$ and the spin.

\subsection{Linear response and superfluid fraction}
\label{sec:linear-response}

As in previous studies, $\bar V_{n,s}$ is assumed to be much smaller than Landau's velocity $V_{n,L}$, and the neutron superfluid density will be evaluated in the limit $\bar V_{n,s}/V_{n,L}\rightarrow 0$. To first order, $\xi_k$ contains terms $\pmb{k}\cdot\pmb{Q}$ therefore $\varepsilon_k$ is independent of $\pmb{Q}$. In uniform neutron matter, $\Delta_k$ is independent of $\pmb{Q}$ for any superfluid velocity $\bar V_{n,s}\leq V_{n,L}$~\cite{allard2023}. In the crust, the dependence of $\Delta_k$ on $\pmb{Q}$ is likely to remain weak therefore 
the first-order correction is neglected. Expanding the neutron mass current~\eqref{eq:average-neutron-mass-current} to first order, taking the continuum limit (the factor of $2$ arises from the spin degeneracy)
\beqy 
\frac{1}{\Omega}\sum_{k} \rightarrow \frac{2}{(2\pi)^3}\int {\rm d}^3\pmb{k} \, ,
\eeqy 
the neutron superfluid density is thus approximately 
\beqy\label{eq:superfluid-density-general} 
\rho_{n,s}=\frac{2}{3}\frac{m_n^2}{\hbar} \int \frac{{\rm d}^3\pmb{k}}{(2\pi)^3}  \vert\V^0_{k\bar{k}}\vert^2 \pmb{\nabla}_{\pmb{Q}}\cdot \pmb{v}_k  \, .
\eeqy 
Here the supercript $0$ refers to quantities in the static ground state. It is understood that the derivatives with respect to $\pmb{Q}$ must be evaluated in the absence of current. The chemical potential $\mu_n \approx \mu^0_n$ is obtained from Eq.~\eqref{eq:average-neutron-mass-density}
\beqy \label{eq:average-neutron-mass-density2}
\bar \rho_n =m_n\frac{2}{(2\pi)^3}\int {\rm d}^3\pmb{k}  \vert\V^0_{k\bar{k}}\vert^2 \, . 
\eeqy 

It is worth remarking that $\pmb{v}_k$ depends on $\pmb{Q}$ explicitly through the combination $\pmb{k}+\pmb{Q}$ from the Floquet-Bloch phase, but also implicitly through the scalar potential $U_n(\rb)$ and the vector potential $\pmb{I_n}(\rb)$ due to the self-consistency of the HFB equations. This can be more clearly seen in the limit of uniform neutron matter at density $\rho_{n,f}$ (empty lattice). 
In this case, the single-particle states are pure plane waves and the single-particle energies are given by~\cite{allard2021}
\beqy
\xi_{k}=\frac{\hbar^2}{2m_n^{\oplus}}\left(\pmb{k}+\pmb{Q}\right)^2 + U_n+\pmb{I_n}\cdot \left(\pmb{k}+\pmb{Q}\right)-\mu_n \, .
\eeqy
Taking the gradient in $\pmb{k}$-space yields 
\beqy 
\pmb{v}_k = \frac{\hbar}{m_n^{\oplus}}\left(\pmb{k}+\pmb{Q}\right) +\frac{\pmb{I_n}}{\hbar} \, .
\eeqy 
Note that the scalar potential $U_n$ does not contribute to $\pmb{v}_k$ in this limiting case. 
The superfluid density can be readily calculated from Eq.~\eqref{eq:superfluid-density-general}: 
\beqy
\rho_{n,s}=\rho_{n,f}\left(\frac{m_n}{m_n^\oplus}+\frac{m_n}{3\hbar^2}\pmb{\nabla}_{\pmb{Q}} \cdot \pmb{I_n} \right)  \, .
\eeqy 
It is only by including the second term that $\rho_{n,s}$ coincides exactly with $\rho_{n,f}$, as explicitly demonstrated in Ref.~\cite{allard2021} (the superfluid density 
was actually calculated for arbitrary temperatures and currents). 
Ignoring the contribution from the vector potential leads to a superfluid fraction $\rho_{n,s}/\rho_{n,f}=m_n/m_n^\oplus$ that differs from 1 unless $m_n^\oplus=m_n$ (in which case $I_n=0$).  Note that the correction $\delta \rho_{n,s}=\rho_{n,f}(1-m_n/m_n^\oplus)$ due to the potential vector is negative  and thus leads to a \emph{reduction} of the superfluid fraction calculated without.

For the inhomogeneous neutron superfluid in the crust, the contribution of the scalar potential $U_n(\rb)$ to $\pmb{v}_k$ no longer cancels. Although $U_n(\rb)$ could implicitly depend on $\pmb{Q}$, the associated leading corrections to the single-particle energies $\epsilon_k$ are necessarily of second order $\pmb{Q}^2$. Therefore these terms will not contribute to $ \pmb{\nabla}_{\pmb{Q}}\cdot \pmb{v}_k $ as in uniform neutron matter recalling that the derivatives here must be evaluated for $Q=0$.  However,  the contribution from the vector potential will not cancel whenever the effective mass differs from the bare mass.  In such case,  neglecting $\pmb{I_n}(\rb)$ as in Refs.~\cite{CCH05a,CCH05b,chamel2006,chamel2012,kashiwaba2019} introduces some inconsistencies. 
This did not have any consequence for the results presented in Refs.~\cite{CCH05a,kashiwaba2019} as the numerical calculations were actually performed for $m_n^\oplus(\rb)=m_n$.  It should be also stressed that the first band-structure calculations predicting a strong suppression of the superfluid fraction in the intermediate layers of the crust were based on models using the bare neutron mass~\cite{chamel2005}.  Treating the neutron superfluid as pure neutron matter, the correction to the superfluid fraction induced by the vector potential is expected to be small since $m_n^\oplus\lesssim m_n$ for the neutron densities $\rho_{n,f}$ prevailing in the inner crust of a neutron star ($\delta  \rho_{n,s}/\rho_{n,f} \approx 3\%$ for the crustal layer considered in this paper, as can be directly seen in Fig.~\ref{fig:effective-mass}).  The role of the vector potential will be further examined in Sec.~\ref{sec:I}. 
Anticipating that $\pmb{I_n}(\rb)$ remains negligible as in uniform neutron matter, $\pmb{v}_k$ depends on $\pmb{Q}$ only through the combination $\pmb{k}+\pmb{Q}$. Therefore, one can write 
\beqy \label{eq:divQv}
\pmb{\nabla}_{\pmb{Q}}\cdot \pmb{v}_k = \pmb{\nabla}_{\pmb{k}}\cdot \pmb{v}_k  \, .
\eeqy 
Integrating Eq.~\eqref{eq:superfluid-density-general} by part, the superfluid density can be calculated as 
\beqy\label{eq:superfluid-density-noI}
\rho_{n,s}=\frac{2}{3}\frac{m_n^2}{\hbar}  \int \frac{{\rm d}^3\pmb{k}}{(2\pi)^3}  \vert\V^0_{k\bar{k}}\vert^2 \pmb{\nabla}_{\pmb{k}}\cdot \pmb{v}_k
=-\frac{2}{3}\frac{m_n^2}{\hbar}  \int \frac{{\rm d}^3\pmb{k}}{(2\pi)^3} \pmb{v}^0_{\pmb{k}} \cdot  \pmb{\nabla}_{\pmb{k}} \vert\V^0_{k\bar{k}}\vert^2\, .
\eeqy 
The surface term has been dropped recalling that the occupation factors $\vert\V^0_{k\bar{k}}\vert^2$ become vanishingly small beyond the Fermi surface, i.e. for energies well above $\mu_n$.  
These occupation factors depend on $\pmb{k}$ through $\varepsilon_{\pmb{k}}$ and $\Delta^0_{\pmb{k}}$. Therefore, 
\beqy\label{eq:grad_Vkk} 
 \pmb{\nabla}_{\pmb{k}} \vert\V^0_{k\bar{k}}\vert^2 =  \hbar \pmb{v}^0_{\pmb{k}} \frac{\partial}{\partial \varepsilon_{\pmb{k}}} \vert\V^0_{k\bar{k}}\vert^2 + \pmb{\nabla}_{\pmb{k}} \Delta^0_{\pmb{k}} \frac{\partial}{\partial \Delta^0_{\pmb{k}}} \vert\V^0_{k\bar{k}}\vert^2 \, .
\eeqy 
In uniform neutron matter, the pairing gaps $\Delta^0_{\pmb{k}}$ are independent of $\pmb{k}$ for the local pairing functionals considered here (see, e.g., Ref.~\cite{allard2021}). 
As shown in Ref.~\cite{chamel2010b}, the pairing gaps in the crust averaged over the Fermi surface deviates at most by about 20\% from the value obtained in uniform 
neutron matter. As can be seen in Fig.~3 of the aforementioned paper, the dispersion of $\D^0_{\pmb{k}}$ for unbound neutrons are of the 
same order. Since the focus of this work is on the superfluid density, $\D^0_{\pmb{k}}=\Delta$ will be treated as a constant as in Refs.~\cite{CCH05b,watanabe2017,minami2022} 
to reduce the computational cost. But different values of $\D$ will be considered. 
Inserting Eq.~\eqref{eq:grad_Vkk}  into Eq.~\eqref{eq:superfluid-density-noI} using Eq.~\eqref{eq:VTerms} yields
\beqy
\rho_{n,s}=\frac{m_n^2}{24 \pi^3 \hbar^2} \int {\rm d}^3\pmb{k} \vert\pmb{\nabla}_{\pmb{k}} \epsilon^0_{\pmb{k}}\vert^2 \frac{\Delta^2}{E_{\pmb{k}}^3} \, , 
\eeqy
where $E_{\pmb{k}}\equiv\sqrt{(\epsilon^0_{\pmb{k}}-\mu_n)^2+\Delta^2}$. 
Let us recall that $\epsilon^0_{\pmb{k}}$ are here evaluated in the absence of current and are therefore periodic functions in $\pmb{k}$-space, i.e., $\epsilon^0_{\pmb{k}+\pmb{G}}=\epsilon^0_{\pmb{k}}$, where $\pmb{G}=m_1 \pmb{b_1}+m_2 \pmb{b_2}+m_3 \pmb{b_3}$ denotes reciprocal lattice vectors, ($m_1$, $m_2$, $m_3$) arbitrary integers and $\pmb{b_1}$, $\pmb{b_2}$, and $\pmb{b_3}$ are the corresponding primitive basis vectors satisfying $\pmb{a_i}\cdot \pmb{b_j}=2\pi \delta_{ij}$ ($\delta_{ij}$ is Kronecker's symbol). Bloch wave vectors are generally imposed to lie within the first Brillouin zone (BZ) by suitable reciprocal lattice translations and by introducing a band index $\alpha$. The volume $\Omega_{\rm BZ}\equiv \vert\pmb{b_1}\cdot \pmb{b_2}\times \pmb{b_3}\vert$ of the first BZ and that of the primitive cell $\Omega_{\rm cell}\equiv \vert\pmb{a_1}\cdot \pmb{a_2}\times \pmb{a_3}\vert$ are related by  $\Omega_{\rm BZ}=(2\pi)^3/\Omega_{\rm cell}$. 
The superfluid density can then be equivalently written as
\beqy\label{eq:superfluid-density-bcs}
\rho_{n,s}=\frac{m_n^2}{24\pi^3 \hbar^2} \sum_{\alpha} \int_{\rm BZ} {\rm d}^3\pmb{k} \vert\pmb{\nabla}_{\pmb{k}} \epsilon^0_{\alpha \pmb{k}}\vert^2 \frac{\Delta^2}{E_{\alpha\pmb{k}}^3}  \,.
\eeqy
This expression is consistent with the mobility coefficient defined as ${\mathcal K}=\rho_{n,s}/m_n^2$ introduced in Ref.~\cite{CCH05b}. The neutron chemical potential $\mu^0_n$ is determined by the mean neutron mass density~\eqref{eq:average-neutron-mass-density}
\beqy\label{eq:chemical-potential}
\bar \rho_n = \frac{m_n}{8\pi^3} \sum_{\alpha}  \int_{\rm BZ} {\rm d}^3\pmb{k} \left(1-  \frac{\epsilon^0_{\alpha\pmb{k}}-\mu^0_n}{E_{\alpha\pmb{k}}}   \right)  \, .
\eeqy 
By symmetry, the $\pmb{k}$-space integration can be further restricted to the  irreducible wedge of the first Brillouin zone (IBZ). For cubic crystals, the integration volume can thus be reduced to $\Omega_{\rm IBZ}=\Omega_{\rm BZ}/48$. 

In the weak coupling limit $\Delta\ll\varepsilon_{\rm F}$ ($\varepsilon_{\rm F}=\hbar^2 k_{\rm F}^2/2m_n$ is the Fermi energy with the Fermi wave number $k_{\rm F}=(3\pi^2 n_{n,f})^{1/3}$), the superfluid density~\eqref{eq:superfluid-density-bcs} reduces to
\begin{eqnarray}\label{eq:super-density-weak1}
	\rho_{n,s} \approx  \frac{m_n^2}{12\pi^3\hbar^2}\sum_\alpha \int_{\rm BZ} d^3\pmb{k}\, |\pmb{\nabla}_{\pmb{k}}\epsilon^0_{\alpha\pmb{k}}|^2 \delta(\epsilon^0_{\alpha\pmb{k}}-\mu^0_n) \, ,
\end{eqnarray}
which can be equivalently written as an integral over the Fermi surface~\cite{CCH05a}
\begin{eqnarray}\label{eq:super-density-weak2}
	\rho_{n,s} = \frac{m_n^2}{12 \pi^3\hbar^2}\sum_\alpha \int_{\rm F} |\pmb{\nabla}_{\pmb{k}} \, 
	\epsilon^0_{\alpha\pmb{k}}|d{\cal S}^{(\alpha)} \, .
\end{eqnarray}
Let us remark that an expression similar to Eq.~\eqref{eq:super-density-weak1} was independently derived in Ref.~\cite{pitaevskii2005} in the context of a dilute 
Fermi gas in a one-dimensional optical potential. 

With the BCS approximation, the neutron superfluid density in neutron-star crusts can thus be extracted semi-analytically from static band-structure calculations using Eq.~\eqref{eq:superfluid-density-bcs}. The  numerical implementation is described in the next section.

\section{Numerical implementation}
\label{sec:numerical-implementation}

For the sake of comparison with previous studies, the same microscopic model as in
 Ref.~\cite{chamel2012} will be employed.  The crust is assumed to be a perfect body-centered cubic lattice crystal 
 of spherical clusters. The primitive basis vectors are given by 
\beqy \label{eq:primitive-vectors}
 \pmb{a_1}&=&\frac{a}{2}\left( -\pmb{\hat x} + \pmb{\hat y} + \pmb{\hat z}\right) \, ,\notag \\ 
 \pmb{a_2}&=&\frac{a}{2}\left( \pmb{\hat x} - \pmb{\hat y} + \pmb{\hat z}\right) \, ,\notag \\ 
 \pmb{a_3}&=&\frac{a}{2}\left( \pmb{\hat x} + \pmb{\hat y} - \pmb{\hat z}\right) \, ,
\eeqy  
where $\pmb{\hat x}$, $\pmb{\hat y}$, $\pmb{\hat z}$ are the Cartesian unit vectors and $a$ is the 
size of the conventional cubic cell. The volume of a primitive cell  
is therefore $\Omega_{\rm cell}=a^3/2$. The reciprocal lattice is a face-centered cubic lattice with primitive basis 
vectors 
\beqy \label{eq:reciprical-primitive-vectors}
 \pmb{b_1}&=&\frac{2\pi}{a}\left(\pmb{\hat y} + \pmb{\hat z}\right) \, ,\notag \\ 
 \pmb{b_2}&=&\frac{2\pi}{a}\left( \pmb{\hat x}  + \pmb{\hat z}\right) \, ,\notag \\ 
 \pmb{b_3}&=&\frac{2\pi}{a}\left( \pmb{\hat x} + \pmb{\hat y} \right) \, .
\eeqy  
 
\begin{figure}
\includegraphics[scale=0.6]{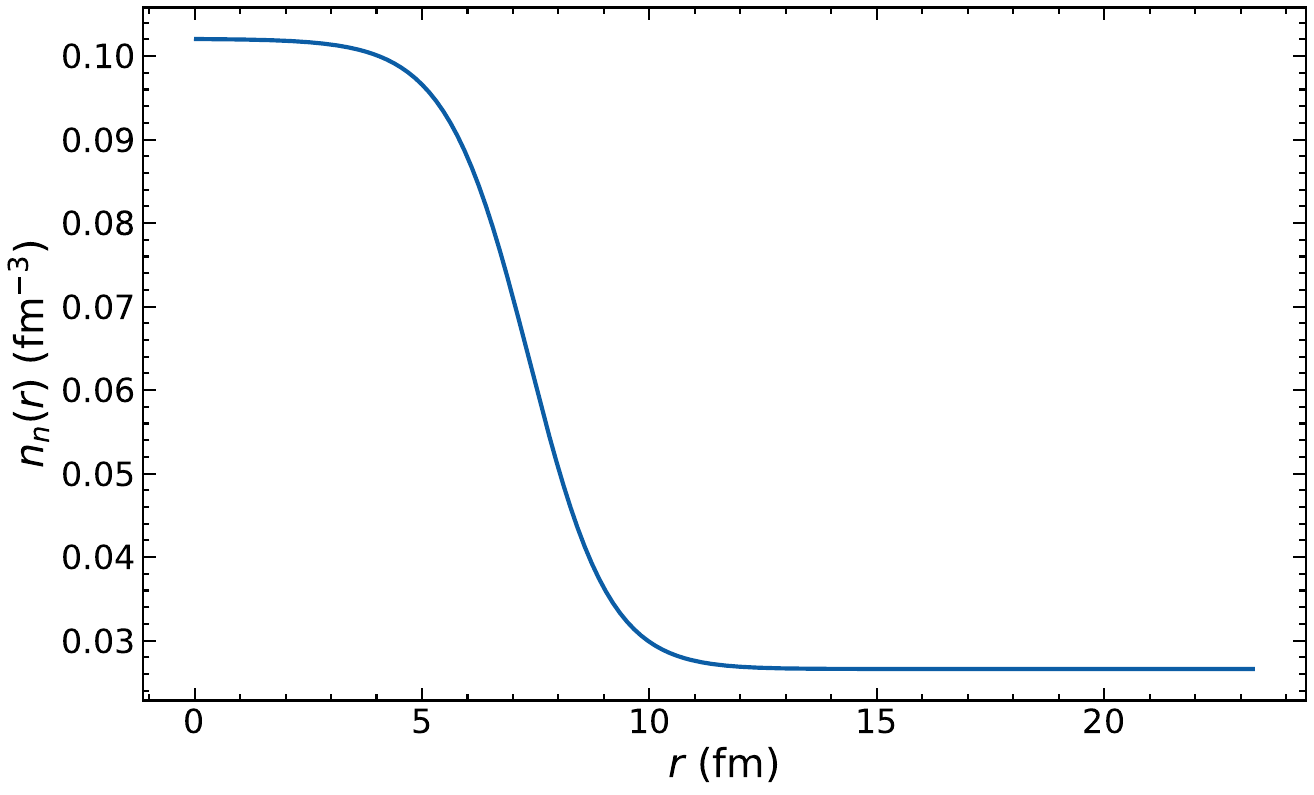}
\vskip -0.5cm
\caption{Neutron density distribution $n_n(r)$ (in fm$^{-3}$)  as function of the radial coordinate $r$ (in fermis) in the Wigner-Seitz cell in the inner crust of a neutron star at average baryon number density 0.03 fm$^{-3}$.}
\label{fig:density}
\end{figure}

\begin{figure}
\includegraphics[scale=0.6]{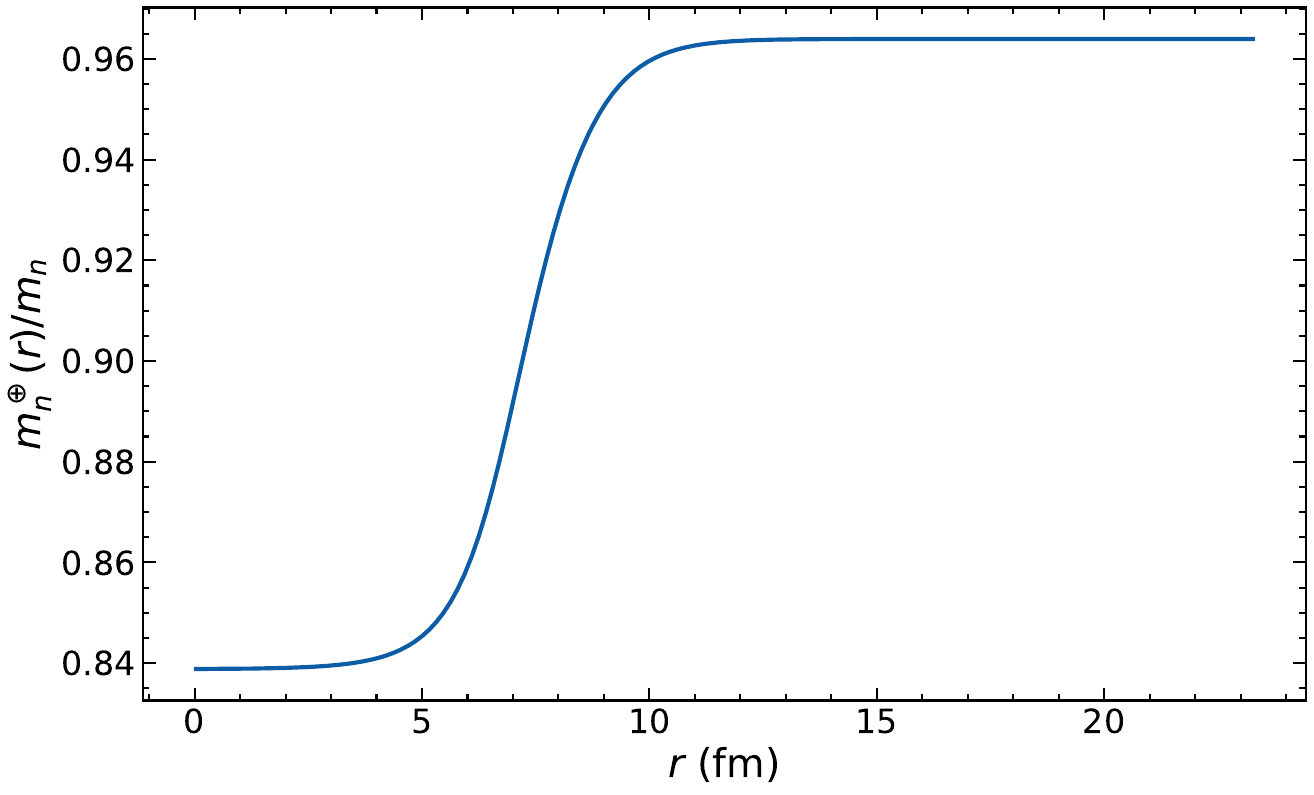}
\vskip -0.5cm
\caption{Local neutron effective mass $m_n^\oplus(r)/m_n$ as function of the radial coordinate $r$ (in fermis) in the Wigner-Seitz cell in the inner crust of a neutron star at average baryon number density 0.03 fm$^{-3}$. }
\label{fig:effective-mass}
\end{figure}

Calculations will be restricted to the crustal layer at average baryon number density 0.03~fm$^{-3}$, 
 where the suppression 
of the neutron superfluid density was found to be the strongest. The Wigner-Seitz cell contains 
$Z_{\rm cell}=40$ protons and $N_{\rm cell}=1550$ neutrons. Its radius is about $R_{\rm cell}=23.30$ fm. 
The lattice spacing $a$ is determined so that the volume\footnote{The Wigner-Seitz cell is a primitive cell, and as such has the same volume as the cell defined by the primitive basis vectors.} $\Omega_{\rm cell}$ of the exact Wigner-Seitz cell (a truncated octahedron) coincides with that 
of the approximate spherical cell, namely $a=R_{\rm cell}(8\pi/3)^{1/3}\approx 47.3$ fm. The neutron number density distribution is plotted in 
Fig.~\ref{fig:density}. The number density of free neutrons estimated from the local neutron density at 
the cell edge is about $n_{n,f}=2.66 \times 10^{-2}$~fm$^{-3}$. This represents about 91\% of the average neutron number density $\bar n_n$. 
The Fermi wave length $\lambda_{\rm F}=2\pi/k_{\rm F}$ of free neutrons %($k_{\rm F}=(3\pi^2 n_{n,f})^{1/3}$ being the Fermi wave number) 
is 
about 6.8 fm. This is much smaller than the lattice spacing and comparable to the radius of the clusters, as can be seen in Fig.~\ref{fig:density}. 
Free neutrons are therefore expected to be scattered not only by the crystal lattice (Bragg diffraction), but also by the individual nucleons 
in the clusters.

In Ref.~\cite{chamel2012}, the neutron superfluid density in the inner crust of a neutron star was evaluated using 
Eq.~\eqref{eq:super-density-weak2} thus ignoring pairing entirely (except for the assumption that a stationary neutron flow in the crust can exist). 
Here pairing will be included using Eq.~\eqref{eq:superfluid-density-bcs}. Varying $\Delta$ from zero to a realistic value will allow us to 
assess the impact of the pairing gap on the superfluid density.

%--------------------------------------------------------------------------------------------------------------------------------
%---------------------------------------------------------------------------------------------------------------------------------

%--------------------------------------------------------------------------------------------------------------------------------
%---------------------------------------------------------------------------------------------------------------------------------

%--------------------------------------------------------------------------------------------------------------------------------
\begin{figure}
\includegraphics[scale=0.6]{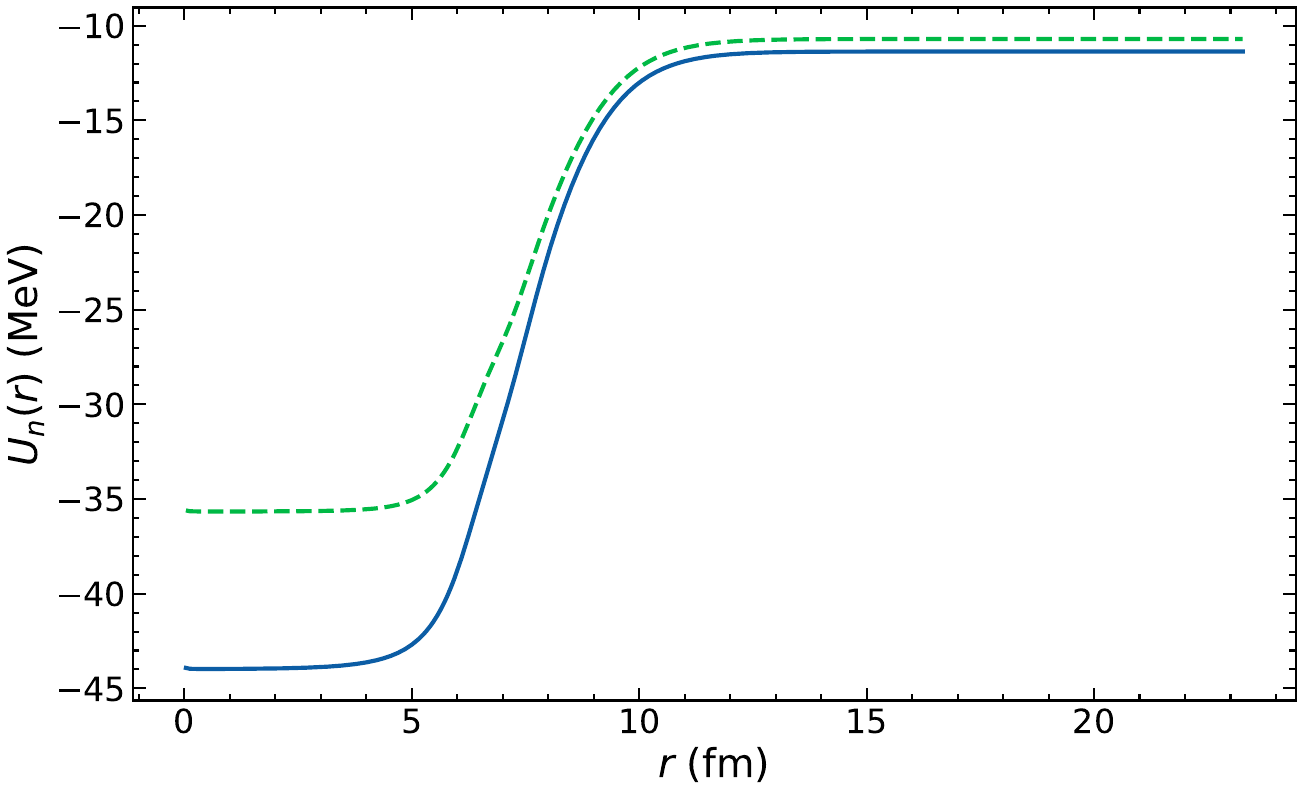}
\vskip -0.5cm
\caption{Local neutron potential $U(r)$ (in megaelectronvolts)  as function of the radial coordinate $r$ (in fermis) in the Wigner-Seitz cell in the inner crust of a neutron star at average baryon number density 0.03 fm$^{-3}$. The dashed line represents the effective potential given by Eq.~\eqref{eq:potential-no-effmass}.}
\label{fig:potential}
\end{figure}
%---------------------------------------------------------------------------------------------------------------------------------

\subsection{Band-structure calculations}
\label{sec:band-structure}

The evaluation of the neutron superfluid density~\eqref{eq:superfluid-density-bcs} requires the determination of the static neutron band structure, i.e., 
the single-particle states satisfying the following Skyrme HF equation:  
\beqy
\label{eq:HF-static}
 -\pmb{\nabla}\cdot \frac{\hbar^2}{2 m_n^\oplus(\rb)}\pmb{\nabla}\varphi_{\alpha\pmb{k}}(\rb,\sigma) + U_n(\rb)\varphi_{\alpha\pmb{k}}(\rb) = \epsilon^0_{\alpha \pmb{k}} \varphi_{\alpha\pmb{k}}(\rb,\sigma) \, ,
\eeqy
with Floquet-Bloch boundary conditions $\varphi_{\alpha\pmb{k}}(\rb+\pmb{\ell},\sigma)=e^{{\rm i} \pmb{k}\cdot \pmb{\ell}}\varphi_{\alpha\pmb{k}}(\rb,\sigma)$
for any lattice translation vector $\pmb{\ell}$.  

The periodic effective mass $m_n^\oplus(\rb)$ and potential $U_n(\rb)$ are constructed by superposing the corresponding mean fields of Ref.~\cite{onsi2008} calculated in the spherical 
Wigner-Seitz cell approximation within the self-consistent 4th-order extended Thomas-Fermi (ETF) method with proton shell correction added consistently via the Strutinsky 
integral (SI) theorem (the correction due to neutron-band structure was shown to be much smaller~\cite{chamel2007} and was therefore neglected). This ETFSI method is a computationally very fast approximation to the HFB equations~\cite{shelley2020}.  In Refs.~\cite{watanabe2017,minami2022}, $m_n^\oplus(\rb)$ was taken as the bare mass $m_n$ and $U_n(\rb)$ was approximated by a fixed sinusoidal potential. The 
realistic fields, however, are quite different, as can be seen in Figs.~\ref{fig:effective-mass} and \ref{fig:potential} respectively, and this can have a 
strong impact on the superfluid density~\cite{chamel2004}. In the present study, the realistic fields will not be approximated. However, they will not be recalculated 
 self-consistently since the fields obtained within the ETFSI approach were shown to be already very close to the self-consistent mean fields~\cite{shelley2020} (see also Fig.2 of Ref.~\cite{pecak2024} for a comparison with full three-dimensional HFB calculations).

The HF equation~\eqref{eq:HF-static} is solved in reciprocal space by expanding the Bloch wavefunctions into Fourier 
series, 
\beqy
\label{eq:plane-wave-expansion}
\varphi_{\pmb{k}}(\rb,\sigma)=\frac{1}{\sqrt{\Omega}}\exp({\rm i}\, \pmb{k}\cdot\rb)\sum_{\pmb{G}} \widetilde{\varphi_{\pmb{k}}}(\pmb{G}) \exp({\rm i} \pmb{G}\cdot\rb)\chi(\sigma)\, ,
\eeqy
where $\chi$ denotes Pauli spinor and the summation is over reciprocal lattice vectors $\pmb{G}$, as imposed by the 
Floquet-Bloch theorem (let us recall that the reciprocal lattice of a body-centered cubic lattice is a face-centered cubic lattice). In this way, Eq.~(\ref{eq:HF-static}) is reduced to a matrix 
eigenvalue problem
\beqy
\label{eq:eigenvalue-problem}
\sum_\beta\biggl[ (\pmb{k}+\pmb{G}_\alpha)\cdot (\pmb{k}+\pmb{G}_\beta) \frac{\hbar^2}{2\widetilde{m}_n^\oplus(\pmb{G}_\beta-\pmb{G}_\alpha)}  + \widetilde{U}_n(\pmb{G}_\beta-\pmb{G}_\alpha)\biggr] \widetilde{\varphi}_{\pmb{k}}(\pmb{G}_\beta) = \epsilon^0_{\alpha\pmb{k}}\, \widetilde{\varphi}_{\pmb{k}}(\pmb{G}_\alpha)
\eeqy
with the Fourier coefficients being expressed by the following integrals over any primitive cell %of volume $V_{\rm cell}$ 
\beqy
\label{24}
\frac{\hbar^2}{2\widetilde{m}_n^\oplus(\pmb{G})}=\frac{1}{\Omega_{\rm cell}}\int d^3\rb\, \frac{\hbar^2}{2m_n^\oplus(\rb)} \exp(-{\rm i}\, \pmb{G}\cdot\rb)\, , \\
\widetilde{U}_n(\pmb{G})=\frac{1}{\Omega_{\rm cell}}\int  d^3\rb\, U_n(\rb) \exp(-{\rm i}\, \pmb{G}\cdot\rb)\, ,
\eeqy 
and with the normalization 
\beqy
\label{eq:norm-wavefunction}
\sum_\beta |\widetilde{\varphi_{\pmb{k}}}(\pmb{G}_\beta)|^2=1\, .
\eeqy

The group velocities $\pmb{v}_{\alpha\pmb{k}} = \pmb{\nabla_k} \epsilon_{\alpha \pmb{k}}/\hbar$ are evaluated using the expression obtained in Ref.~\cite{ChamelAllard2019}~: 
\beqy
\label{eq:group-velocity}
\pmb{v}_{\alpha\pmb{k}}=\sum_\sigma\int d^3\rb\,  \varphi_{\alpha\pmb{k}}(\rb,\sigma)^* \pmb{v}(\rb)\varphi_{\alpha\pmb{k}}(\rb,\sigma) 
\eeqy
where the velocity operator is defined as 
\beqy
\pmb{v}(\rb) =\frac{1}{\mathrm{i} \hbar }\biggl[\rb h_n(\rb) - h_n(\rb)\rb \biggr]=\frac{-{\rm i} \hbar}{2}\left(\frac{1}{m_n^\oplus(\rb)}\pmb{\nabla}+\pmb{\nabla}\frac{1}{m_n^\oplus(\rb)}\right)\, . 
\eeqy
The expression~\eqref{eq:group-velocity} can be efficiently computed using fast Fourier transforms (FFT). 
Equation~\eqref{eq:group-velocity} has been checked numerically by computing the derivatives in $\pmb{k}$-space using finite differences. The results are found to be in excellent agreement.

In numerical implementations, summations are calculated keeping a finite number of terms, $\mathcal{N}_1$, $\mathcal{N}_2$, and $\mathcal{N}_3$ along each basis vectors in real space or reciprocal space. FFT are used to switch representations between real and reciprocal spaces. 
Since the lattice has cubic symmetry, one can choose $\mathcal{N}_1=\mathcal{N}_2=\mathcal{N}_3\equiv \mathcal{N}$. The number of terms to include in the expansion cannot be arbitrarily small, but is dictated by the number of neutrons in the layer of interest. Indeed, each energy band can accommodate two neutrons (with opposite spins) per primitive cell at most (less if neutrons are paired due to the smearing of the Fermi-Dirac distribution). Therefore, $N_{\rm cell}$ neutrons per cell occupy $N_{\rm cell}/2$ bands at least. However, the maximum number of bands of the discrete eigenvalue problem~\eqref{eq:eigenvalue-problem} is limited by $\mathcal{N}^3$. In other words, $\mathcal{N}$ should be greater than $(N_{\rm cell}/2)^{1/3}$ to ensure that no significant violation of neutron number conservation will arise (pairing further increases the minimum number of terms although the shift is not expected to be large since most bands will still accommodate the same number of neutrons except for those lying around the Fermi level). 
Here $N_{\rm cell}=1550$ so that one should set $\mathcal{N}\geq 10$. This represents an absolute lower limit. However, much finer grids are needed to compute the single-particle energies at the required accuracy. 

Truncating the Fourier expansion implies a discretization of the real space. Namely, the HF equations are solved in a spatial grid inside the rhombohedral primitive cell of the body-centered cubic lattice defined by the primitive basis vectors. The grid points are defined by $\pmb{r}=(i_1/\mathcal{N})\pmb{a_1}+(i_2/\mathcal{N})\pmb{a_2}+(i_3/\mathcal{N})\pmb{a_3}$, where $i_1$, $i_2$, and $i_3$ are positive integers less or equal than $\mathcal{N}-1$. Using Eq.~\eqref{eq:primitive-vectors}, this corresponds to the following Cartesian coordinates: 
\beqy 
x&=&\frac{a}{2\mathcal N}\left( -i_1 +i_2 + i_3\right) \, ,\notag \\
y&=&\frac{a}{2\mathcal N}\left( i_1 -i_2 + i_3\right) \, ,\notag \\
z&=&\frac{a}{2\mathcal N}\left( i_1 +i_2 - i_3\right) \, .
\eeqy 
The spacing along each Cartesian axis of the conventional cubic cell is therefore $\delta x=\delta y=\delta z= a/(2 \mathcal{N})$  (the spacing along each primitive axis is $\sqrt{3}$ times larger). 
In other words, a grid spacing $\delta x$ thus requires $\sim (a/2\delta x)^3$ points. 
Here $a=47.3$~fm. 
The size of the clusters is about an order of magnitude smaller, as can be seen in Figs.~\ref{fig:effective-mass} and \ref{fig:potential}. 
The number of terms in the plane-wave expansion should be large enough to properly describe the internal structure of clusters since the wavefunctions of unbound neutrons must still be orthogonal to those of bound states. Unless stated otherwise, calculations have been performed using a grid of $25\times 25\times 25 = 15625$ points corresponding to a grid spacing $\delta x \approx 0.95$~fm (about 1.65 fm along each primitive axis). 
For comparison, three-dimensional mean-field calculations of neutron-star crusts are typically carried out in a uniform Cartesian grid with a spacing $\delta x=\delta y=\delta z\approx 1-1.3$ fm (see, e.g., Refs.~\cite{Magierski2002,Gogelein2007,Schuetrumpf2019,Newton2022}). 

%spatial resolution employed in previous studies: 
% 1.3 fm in Magierski & Heenen (2002)
% 1 fm in Gogelein & Muether (2007)
% around > 1 fm in Schuetrumpf et al. (2019)
% 1.2-1.3 fm in Newton et al. (2022)

In the absence of pairing, each band can accommodate two neutrons (with opposite spins) per Wigner-Seitz cell at most.  %(less when pairing is included) 
Given the number $N_{\rm cell}$ of neutrons per Wigner-Seitz cell, the number of bands should thus be larger than $N_{\rm cell}/2=775$ in the present case. If neutrons are paired, the occupancy of each band is reduced  due to the smearing of the Fermi-Dirac distribution and in principle summations have to be performed over all bands. However, the factor $\D^2/E^3$ in the integral for the superfluid fraction~\eqref{eq:superfluid-density-bcs} falls rapidly whenever $\xi$ departs from $\xi=0$. For $\xi_{\rm max}\approx  4.53 \D$, the factor 
$\D^2/E^3$ is reduced by a hundred compared to its maximum value for $\xi=0$. In actual numerical calculations, the contribution from Bloch states with energies $\epsilon^0_{\alpha\pmb{k}}>\epsilon_{\rm max}\equiv \xi_{\rm max}+\mu^0_n $ will thus be ignored. Since the density of unbound states is given to a very good approximation by that of an ideal Fermi gas~\cite{chamel2009b}, the number of bands up to $\epsilon_{\rm max}$ is roughly given by 
\beqy\label{eq:max-bands}
\alpha_{\rm max}\equiv \alpha (\epsilon_{\rm max}) \approx Z_{\rm cell} + \frac{\Omega_{\rm cell}}{6 \pi^2} \left[\frac{2m^\oplus_B (\epsilon_{\rm max}-U_{ B})}{\hbar^2}\right]^{3/2}\, , 
\eeqy 
where the number of bands corresponding to bound states is roughly estimated as the number of protons per Wigner-Seitz cell (in other words, the number of bound neutrons is taken to be twice that of protons), $m_{B}^\oplus$ and $U_{B}$ are the mean values of the Skyrme effective mass $m_n^\oplus(\rb)$ and  potential $U_n(\rb)$ in the interstitial ``background'' region between clusters respectively. Since the effective mass is generally smaller than the bare mass, as can be seen in Fig.~\ref{fig:effective-mass}, an upper limit can be obtained by setting $m_B^\oplus=m_n$.  
Substituting $\epsilon_{\rm max}=4.53 \D + \mu^0_n$ with the realistic value $\Delta=1.59$~MeV for the pairing gap (as obtained from diagrammatic calculations in neutron matter at the same average neutron density taking into account both polarization and self-energy effects~\cite{cao2006}) and estimating the chemical potential $\mu^0_n$ using Eq.~(B21) of Ref.~\cite{pearson2012} yield $\alpha_{\rm max} \approx 1265$. This rough estimate has been confirmed by extensive numerical calculations. 
To better assess the sensitivity of the superfluid fraction with respect to the pairing gap, lower values of $\Delta$ as well down to the limit 
$\Delta=0$ will be considered.

\subsection{Integrations over the first Brillouin zone}
\label{sec:BZ-integrations}

Integrations over the first Brillouin zone have been carried out using the special-point method~\cite{chadi1973}. The superfluid fraction integral~\eqref{eq:superfluid-density-bcs} is replaced by a discrete sum 
over a set of points uniquely determined by the crystal structure
\begin{equation}\label{eq:superfluid-density-bcs-special}
	\rho_s\approx \frac{2 m_n^2}{\pi^3\hbar^2} \Omega_{\rm IBZ}\sum_{p=1}^{p_{\rm max}} w_p \sum_{\alpha=1}^{\alpha_{\rm max}}  |\pmb{\nabla}_{\pmb{k_p}}
	\epsilon^0_{\alpha\pmb{k_p}}|^2 \frac{|\Delta|^2}{(E_{\alpha\pmb{k_p}})^3}\, .
\end{equation}
The formulae given in Ref.~\cite{hama1992} for the components of the wave vectors $\pmb{k_p}$ and the associated weights $w_p$ have been used. This method assumes that the integrand is sufficiently smooth such that the (symmetrized) Fourier components of the integrand decrease rapidly with increasing norm of the reciprocal lattice vector. 

For very small gaps $\D$, the factor  $\D^2/E^3$ is very strongly peaked around $\xi=0$ so that 
the special-point method becomes less reliable. In the absence of pairing $\D=0$, the special-point 
method can still provide fairly accurate results by smearing the Fermi-Dirac distribution so as to 
remove the discontinuity at the Fermi surface. 
A popular method originally developed for electronic-structure calculations is that of Methfessel and Paxton~\cite{methfessel1989} based on an approximate representation of the Dirac delta distribution in terms of Hermite polynomials $H_q(x)$ of order $q$ ($q$ being an integer), namely 
\beqy  \label{eq:MethfesselPaxton}
\delta(x)\approx D(x) \equiv 1 + \frac{\exp(-x^2)}{\sqrt{\pi}}\sum_{q=1}^{q_{\rm max}} \left(\frac{-1}{4}\right)^q \frac{H_{2q}(x)}{q!}\, ,
\eeqy 
with $x=(\epsilon - \mu^0_n)/W$. The smearing width $W$ is a free parameter that must be suitably 
adjusted together with the integer $q_{\rm max}$. By construction, for any polynomial $P(x)$ of order $2q_{\rm max}+1$ or less, one  has 
\beqy 
\int_{-\infty}^{+\infty} dx  D(x) P(x) = \int_{-\infty}^{+\infty} dx \delta (x) P(x)=P(0)\, .
\eeqy 
Smearing the Fermi-Dirac distribution improves the convergence of the special-point method but requires the computation of additional bands lying above the 
Fermi level. The minimum number of bands to consider can be estimated similarly to Eq.~\eqref{eq:max-bands} with $\epsilon_{\rm max}\approx \mu^0_n + W$. 
Benchmark calculations of the lattice Green functions are presented in the Appendix.

\subsection{Integrations over the Fermi surface}
\label{sec:FS-integrations}

In the limit of very weak pairing, the special point method may become unreliable for evaluating Eq.~\eqref{eq:superfluid-density-bcs} due to the very sharp variations of $\D^2/E^3$ with $\xi$. The superfluid fraction in the limiting case $\D/\epsilon_{\rm F}\rightarrow 0$ has been computed using Eq.~\eqref{eq:super-density-weak2}. Such integrations can be accurately evaluated using the semianalytical method of Gilat and Raubenheimer~\cite{gilat1966}. 
The irreducible domain is decomposed into identical cubic cells, whose centers are located at Bloch wave vectors $\pmb{k_c}$ with Cartesian coordinates $k_{cx}=(2j_1+1)\pi/(a\mathcal{N}_c)$, $k_{cy}=(2j_2+1)\pi/(a\mathcal{N}_c)$ and $k_{cz}=(2j_3+1)\pi/(a\mathcal{N}_c)$, where $j_1$, $j_2$, and $j_3$ are integers satisfying $\mathcal{N}_c-1\geq j_1\geq j_2\geq j_3\geq 0$ and $j_1+j_2\leq \mathcal{N}_c-1$. Here $\mathcal{N}_c$ denotes the number of cells along the axis $k_x$. The total number of cubic cells in the irreducible domain is $c_{\rm max}=\mathcal{N}_c(\mathcal{N}_c+2)(\mathcal{N}_c+4)/24$ for $\mathcal{N}_c$ even and $(\mathcal{N}_c+1)(\mathcal{N}_c+2)(\mathcal{N}_c+3)/24$ for $\mathcal{N}_c$ odd. 
By approximating the Fermi surface intersections inside each cell by a plane, the contributions of each cell to the integral can be evaluated analytically. The superfluid fraction integral~\eqref{eq:super-density-weak2} is thus replaced by a discrete sum over cubic cells  
\beqy\label{eq:super-density-weak2-GR}
	\rho_s \approx \frac{m_n^2}{12\pi^3\hbar^2}\sum_{c=1}^{c_{\rm max}} w_c \sum_\alpha  |\pmb{\nabla}_{\pmb{k_c}} \, 
	\epsilon^0_{\alpha\pmb{k_c}}| {\cal S}_c^{(\alpha)} \, ,
\eeqy
where 
 $\mathcal{S}_c^{(\alpha)}$ represents the area of the branch of Fermi surface from the band $\alpha$ inside the cell under consideration. 
Expressions for $\mathcal{S}_c^{(\alpha)}$ can be found in Ref.~\cite{gilat1966}. The introduction of the weights $w_c$ is to account for the fact that some of the cells lie outside the irreducible domain. Their determination is discussed in Ref.~\cite{janak1971}.

In principle, the evaluation of Eq.~\eqref{eq:super-density-weak2-GR} only requires the computation of the energies $\epsilon^0_{\alpha\pmb{k_c}}$ in the vicinity of the Fermi level, more precisely $\vert \mu^0_n - \epsilon^0_{\alpha\pmb{k_c}}\vert  \leq \sqrt{3}\,\vert \pmb{\nabla}_{\pmb{k_c}} \epsilon^0_{\alpha\pmb{k_c}}\vert \pi/(a \mathcal{N}_c) $. However,  the gradients $\vert \pmb{\nabla}_{\pmb{k_c}} \epsilon^0_{\alpha\pmb{k_c}}\vert$ are difficult to reliably estimate a priori unlike the chemical potential $\mu^0_n$. With the approximation $\vert \pmb{\nabla}_{\pmb{k_c}} \epsilon^0_{\alpha\pmb{k_c}}\vert \approx  \hbar^2 k_{\rm F}/m_B^\oplus$, % where $k_{\rm F}$ is the Fermi wave vector of unbound neutrons, 
the minimum number of bands to consider can still be reasonably well determined from Eq.~\eqref{eq:max-bands} by substituting $\epsilon_{\rm max}=\mu^0_n+\sqrt{3} \pi  \hbar^2 k_{\rm F}/(a \mathcal{N}_c m_B^\oplus) $. Computations could be further reduced by ignoring bands with energies below $\epsilon_{\rm min}=\mu^0_n-\sqrt{3} \pi  \hbar^2 k_{\rm F}/(a \mathcal{N}_c m_B^\oplus) $. However, the results obtained in this way have been found to become less accurate with increasing number of cells. This stems from the fact that as the energy window around $\mu^0_n$ shrinks, the results become more sensitive to the errors in our rough estimate of $\epsilon_{\rm min}$ and  $\epsilon_{\rm max}$.  
To avoid introducing errors that remain difficult to control, all bands up to $\epsilon_{\rm max}$ have been included even though the calculations are computationally more expensive.  
Benchmark calculations are presented in the Appendix.

\section{Results and discussion}
\label{sec:results}

\subsection{Weak-coupling approximation}
\label{sec:weak-coupling}

\begin{figure}
	\includegraphics[scale=0.6]{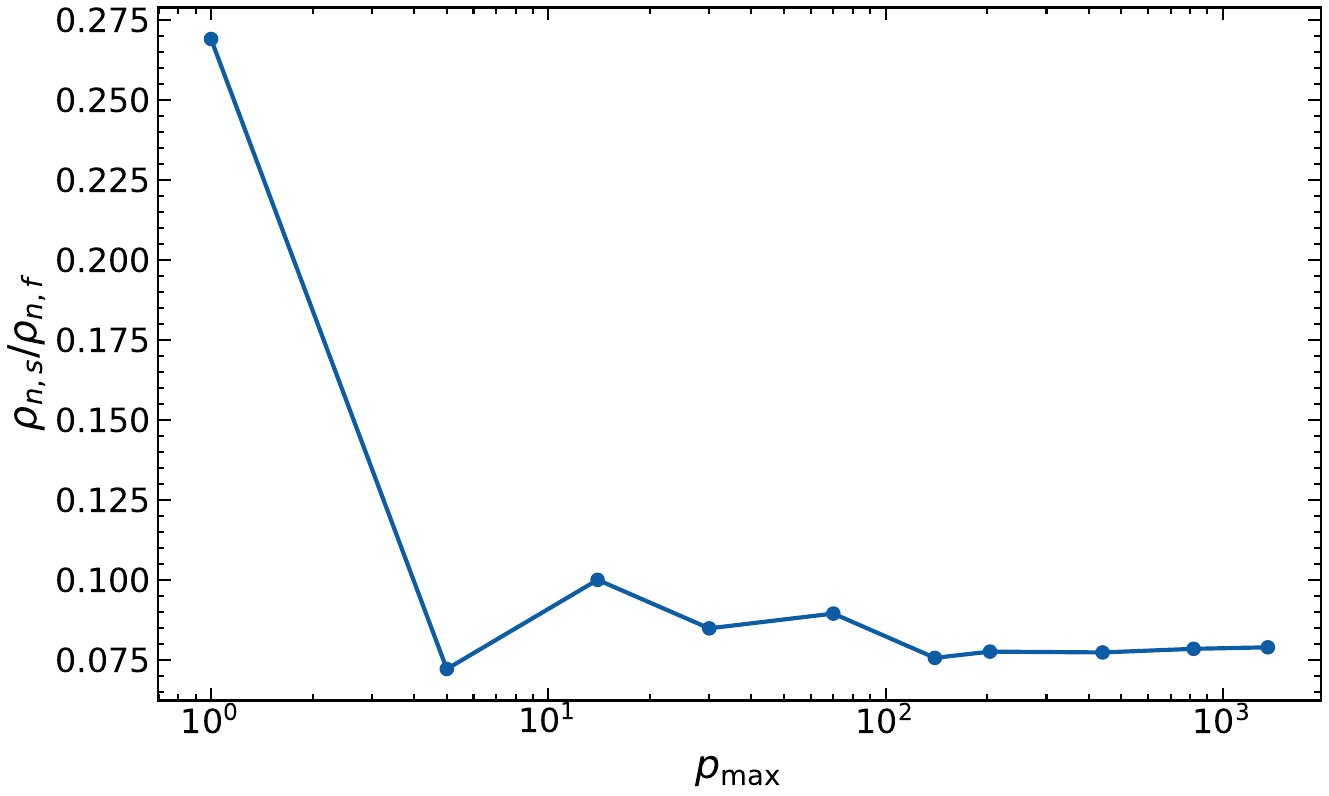}
	\vskip -0.5cm
	\caption{Convergence of the neutron superfluid fraction as calculated from Eq.~\eqref{eq:super-density-weak1}  with respect to the number of special points. }
	\label{fig:convergence-special-superfluid-fraction-nopairing}
\end{figure}

The neutron superfluid density has been first calculated in the limit $\D/\epsilon_{\rm F} \rightarrow 0$ from Eq.~\eqref{eq:super-density-weak1}.  Brillouin-zone integrations have been performed using the special-point method with $p_{\rm max}=1360$, i.e. 65280 wave vectors in the first Brillouin zone. The Fermi-Dirac distribution has been approximated by the method of Methfessel and Paxton with $30$ Hermite polynomials varying the smearing width $W$ from 0.02 MeV to 2 MeV. The number of polynomials has been increased up to $40$ but the results have not changed appreciably. About 900 bands have been included. The chemical potential has been calculated using Eq.~(B21) of Ref.~\cite{pearson2012}.  
Results are summarized in Table~\ref{tab1}. For comparison with previous studies, the superfluid density is expressed in units of the density of free neutrons. Averaging over $W$ yields $\rho_{n,s}/\rho_{n,f} \approx 7.90\%$. Even though $W$ varies by two orders of magnitude, the results remain remarkably stable; the deviations compared to the average value do not exceed 3\%.  The convergence with respect to the number of special points is shown in Fig.~\ref{fig:convergence-special-superfluid-fraction-nopairing}.

\begin{figure}
	\includegraphics[scale=0.6]{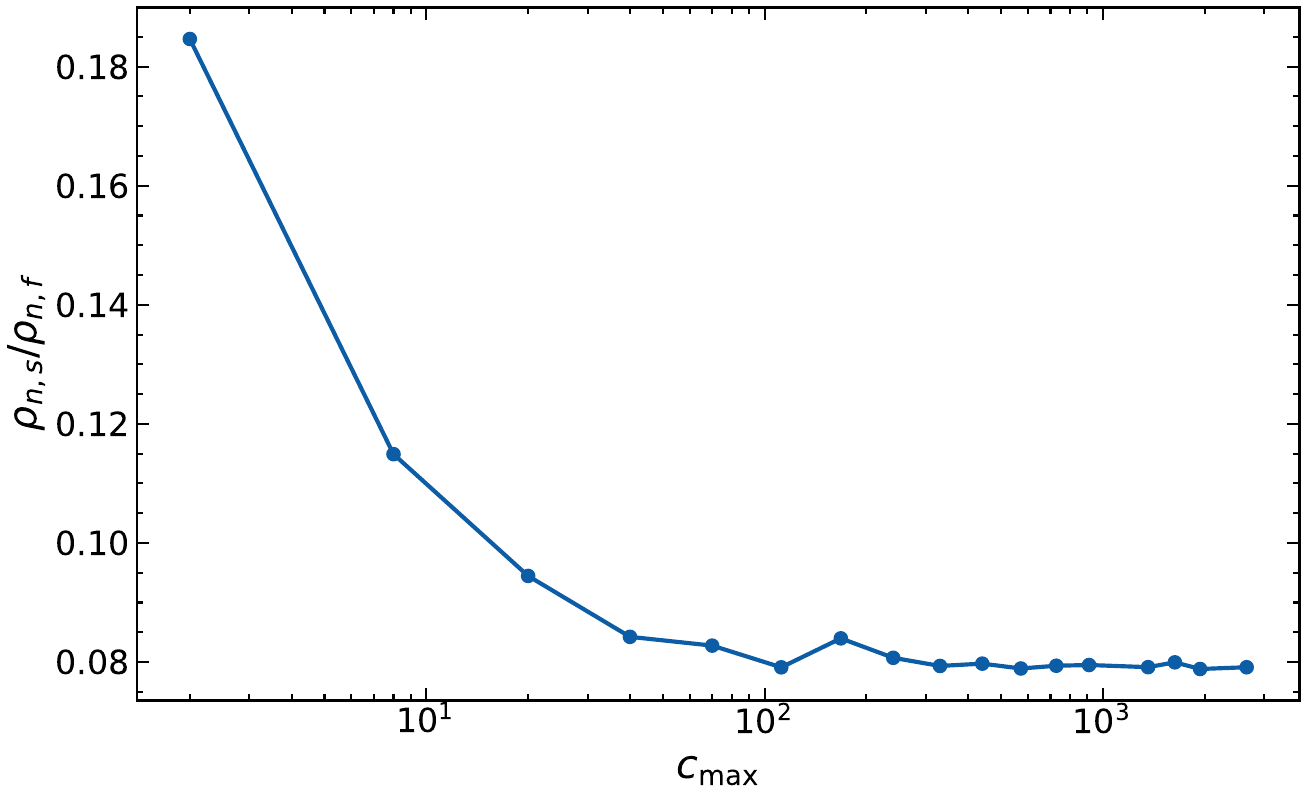}
	\vskip -0.5cm
	\caption{Convergence of the neutron superfluid fraction as calculated from Eq.~\eqref{eq:super-density-weak2}  with respect to the number of cubic cells. }
	\label{fig:convergence-GR-superfluid-fraction}
\end{figure}

\begin{figure}
	\includegraphics[scale=0.6]{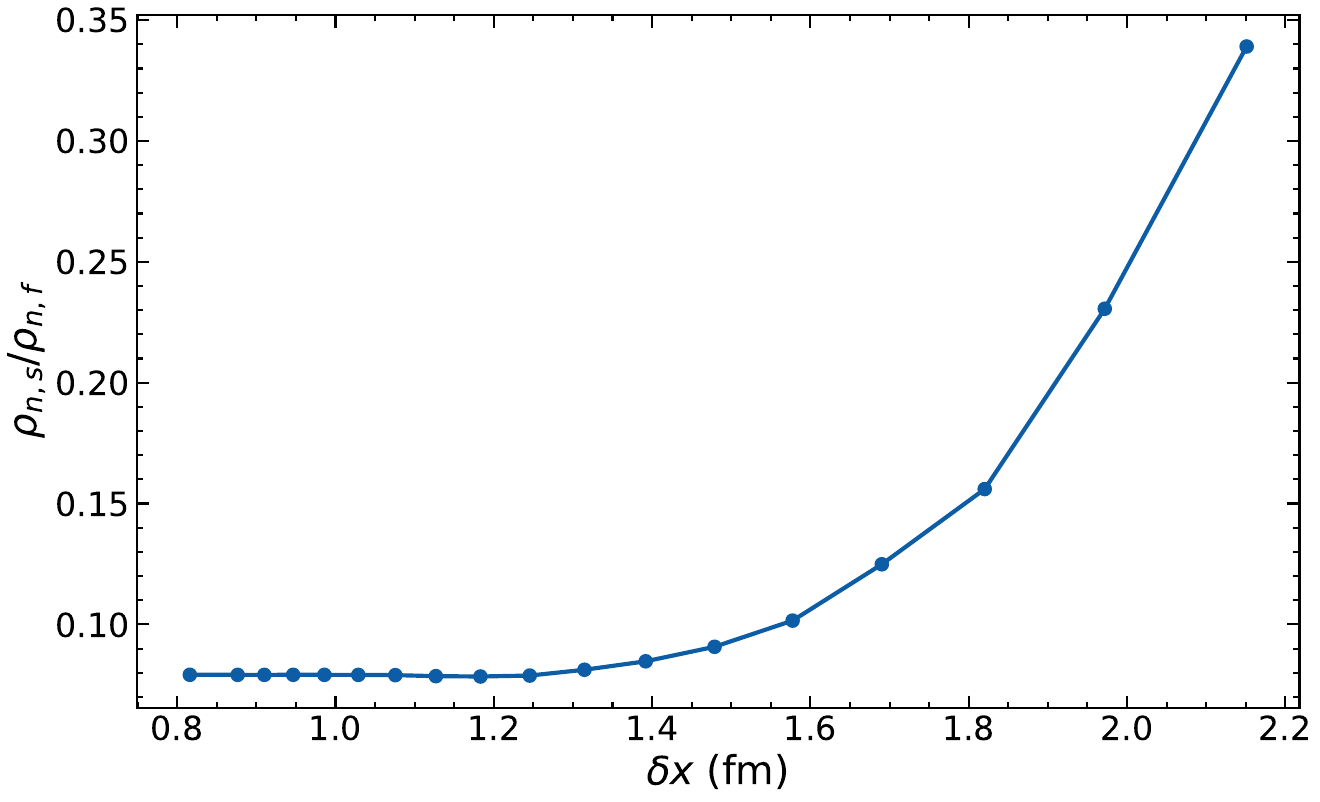}
	\vskip -0.5cm
	\caption{Convergence of the superfluid fraction with respect to the spatial grid spacing $\delta x=\delta y=\delta z$ (in fermis) along each Cartesian axis of the conventional cubic cell (the spacing along each primitive axis is $\sqrt{3}$ times larger). }
	\label{fig:convergence-resolution}
\end{figure}

For comparison, Fermi-surface integrations of Eq.~\eqref{eq:super-density-weak2} using the more accurate Gilat-Raubenheimer method with up to 2660 cubic cells in the irreducible domain (127680 in the first Brillouin zone) including about 800 bands lead to $\rho_{n,s}/\rho_{n,f}\approx 7.91\%$, as shown in Fig.~\ref{fig:convergence-GR-superfluid-fraction}. The results obtained with these two completely different methods are thus found to be in excellent agreement. The Gilat-Raubenheimer method, however, is computationally less demanding and is to be preferred. This value also agrees within 7\% with that obtained in Ref.~\cite{chamel2012} using a different code. 
The reliability of these results has been tested by recalculating the chemical potential using the special-point method. The chemical potential is not very sensitive to the 
detailed band structure~\cite{chamel2007}. The convergence of the superfluid fraction up to 3 significant digits has been achieved using only 70 special points in the irreducible domain (3360 wave vectors in the first Brillouin zone). The result is found to be the same as that obtained previously.
The convergence with respect to the grid spacing has been also tested. Results are plotted in Fig.~\ref{fig:convergence-resolution}. The superfluid fraction is calculated using the Gilat-Raubenheimer method with $c_{\rm max}=1360$ (65280 cubic cells in the first Brillouin zone). The highest (lowest) resolution is reached with a FFT grid of $29\times 29\times 29=24389$ ($10\times 10\times 10=1000$) points.  Fairly accurate results can be obtained with $\delta x \approx 1.3$~fm (2.3 fm along each primitive axis) corresponding to a FFT grid of $18\times 18\times 18=5832$ points. Increasing $\delta x$ means reducing the number of plane waves describing each Bloch state.  
The scattering of free neutrons is therefore artificially suppressed and the superfluid fraction is enhanced. For the lowest resolution with $\delta x\approx 2.2$~fm (3.8 fm along each primitive axis), 
the superfluid fraction deviates from the exact result by a factor $4.3$. The superfluid fraction approximately increases with the grid spacing as a power law: $\rho_{n,s}/\rho_{n,f}\approx 0.076 + 0.00119 \delta x^{7.07}$.

\begin{table}
	\centering
	\caption{Neutron superfluid fraction in the inner crust of a neutron star calculated from Eq.~\eqref{eq:super-density-weak1} using 1360 special points in the irreducible domain and the smeared Fermi-Dirac distribution of Methfessel and Paxton with $30$ Hermite polynomials and different smearing widths $W$.} 
	\label{tab1}
	\begin{tabular} {|c|c|}
	\hline 
     $W$ (MeV) & $\rho_{n,s}/\rho_{n,f}$ \\ 
     \hline 
	0.020000 & 0.075727\\
	0.025485 & 0.077381\\
	0.032476 & 0.078043\\
	0.041383 & 0.075969\\
	0.052733 & 0.076040\\
	0.067196 & 0.077659\\
	0.085627 & 0.077774\\
	0.109112 & 0.078636\\
	0.139039 & 0.079420\\
	0.177173 & 0.078639 \\
	0.225768 & 0.078649\\
	0.287690 & 0.078712 \\
	0.366596 & 0.078586\\
	0.467144 & 0.078431\\
	0.595270 & 0.078523\\
	0.758538 & 0.078660\\
	0.966586 & 0.078207\\
	1.231696 & 0.078290\\
	1.569520 & 0.077936\\
	2.000000 & 0.077097\\
		\hline
	\end{tabular}
\end{table}

%--------------------------------------------------------------------------------------------------------------------------------
%---------------------------------------------------------------------------------------------------------------------------------

%--------------------------------------------------------------------------------------------------------------------------------
%---------------------------------------------------------------------------------------------------------------------------------

%--------------------------------------------------------------------------------------------------------------------------------
%---------------------------------------------------------------------------------------------------------------------------------

%--------------------------------------------------------------------------------------------------------------------------------
\begin{figure}
	\includegraphics[scale=0.6]{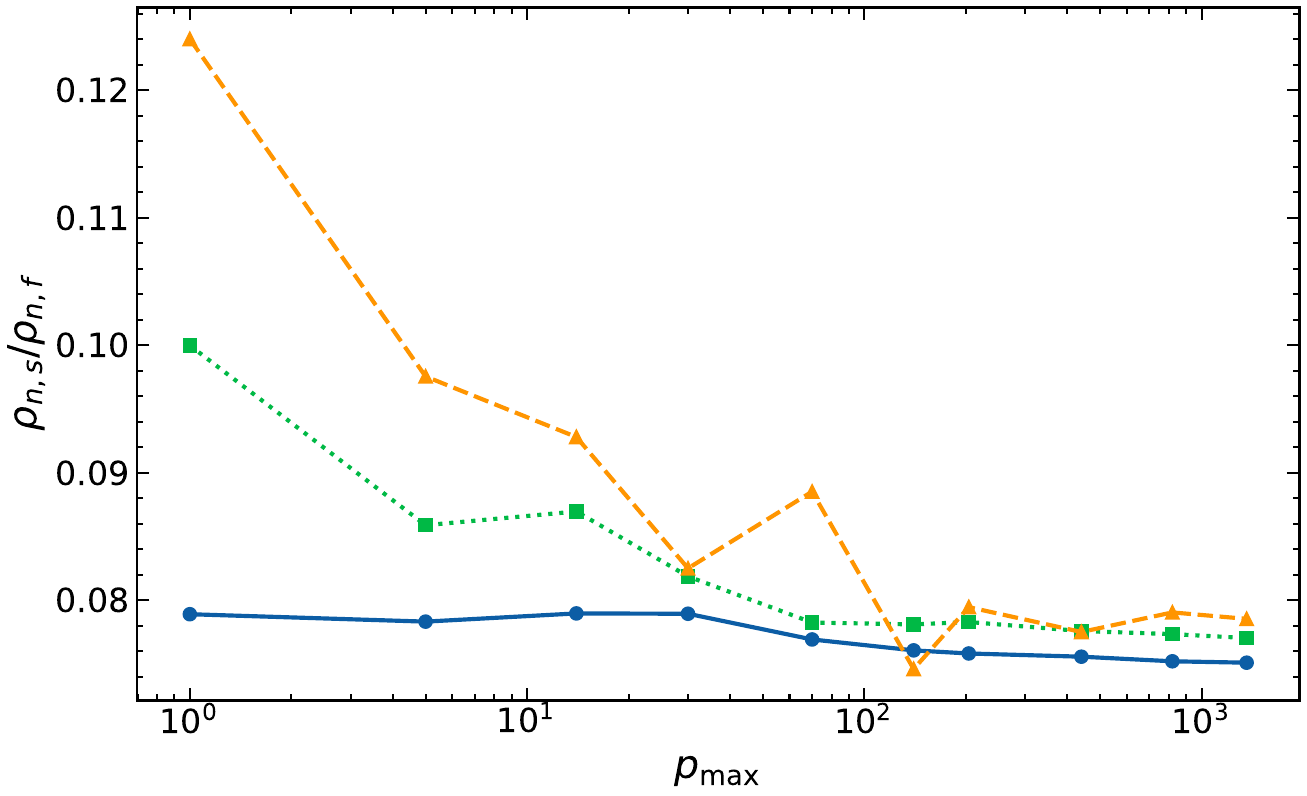}
	\vskip -0.5cm
	\caption{Convergence of the neutron superfluid fraction as calculated from Eq.~\eqref{eq:superfluid-density-bcs}  with respect to the number of special points. The solid curve is for $\Delta=1.59$~MeV, the dotted curve for $\Delta=0.180$~MeV, and the dashed curve for $\Delta=0.0203$~MeV.}
	\label{fig:convergence-special-superfluid-fraction-pairing}
\end{figure}
%---------------------------------------------------------------------------------------------------------------------------------

\subsection{Role of BCS pairing}
\label{sec:BCS}

Having assessed the precision of the special-point method for computing the superfluid fraction, Eq.~\eqref{eq:superfluid-density-bcs-special} has been evaluated with $p_{\rm max}=1360$ (65280 wave vectors in the first Brillouin zone) for different pairing gaps using an FFT grid of $25\times 25\times 25$ points. These calculations are computationally more demanding due to the much higher number of bands to consider, about 1300. This means computing about $2\times 10^7$ Bloch single-particle states. As discussed above and as can be seen in Fig.~\ref{fig:convergence-special-superfluid-fraction-pairing}, the larger $\Delta$, the faster is the convergence and the more reliable are the results. For the realistic value of $\Delta=1.59$~MeV, the value of $\rho_{n,s}/\rho_{n,f}$ obtained using the mean-value point agrees within 5\% with the value obtained using 1360 special points. The superfluid fractions calculated for different pairing gaps are summarized in Table~\ref{tab1}. The relative deviations do not exceed 7\% and lie within the precision of the numerical methods. In other words, the variations of superfluid fraction with the pairing gap are not significant, thus confirming earlier expectations~\cite{CCH05b}. Recalculating the chemical potential from Eq.~\eqref{eq:chemical-potential} using up to $p_{\rm max}=70$ changes the numerical results for the superfluid fraction by about 1\% at most for the lowest values of $\Delta$. The computational cost of the superfluid fraction can thus be substantially reduced by ignoring pairing.

\begin{table}
	\centering
	\caption{Neutron superfluid fraction in the inner crust of a neutron star for different pairing gaps $\Delta$ (in megaelectronvolts).} 
	\label{tab2}
	\begin{tabular} {|c|c|}
		\hline 
		$\D$ (MeV) & $\rho_{n,s}/\rho_{n,f}$ (\%) \\ 
		\hline 
		1.59 &  7.51\\ 
		1.11 &  7.51\\ 
		0.770 &  7.52\\ 
		0.535 &  7.55\\ 
		0.372 &  7.60\\ 
		0.259 &  7.66\\ 
		0.180 &  7.70\\ 
		0.125 &  7.74\\ 
		0.0869 &  7.77\\ 
		0.0604 &  7.79 \\
		0.0420 & 7.78 \\ 
		0.0292 & 7.78 \\ 
		0.0203 & 7.79 \\ 
		0.0141 & 7.89 \\ 
		0.00981 & 7.93 \\ 
		\hline
	\end{tabular}
\end{table}

\subsection{Role of the effective mass and the vector potential}
\label{sec:I}

In the calculations presented so far, the vector potential $\pmb{I_n}(\rb)$ was ignored thus introducing some inconsistencies, as previously discussed in Sec.~\ref{sec:linear-response}.  The associated errors have been estimated as follows. The vector potential $\pmb{I_n}(\rb)$ arises in the presence of currents whenever the single-particle Hamiltonian contains an effective mass $m_n^\oplus(\rb)$.
In turn, the deviation of this effective mass from the bare mass originates from terms depending on the kinetic density $\tau_n(\rb)$ in the Skyrme potential energy $E_{\rm pot}$,  see Eq.~\eqref{eq:Fields1}.  In those
terms, the Thomas-Fermi expression is used, namely (see, e.g., Ref.~\cite{Brack_ea85})
\beqy 
\tau_n(\rb)\approx \frac{3}{5}(3\pi^2)^{2/3} n_n(\rb)^{5/3} \, . 
\eeqy 
The kinetic density $\tau_n(\rb)$ is kept explicitly only in the kinetic energy
\beqy 
E_{\rm kin}=\sum_q \int {\rm d}^3\rb\, \frac{\hbar^2}{2m_q} \tau_q(\rb) \, .
\eeqy 
Having eliminated $\tau_n(\rb)$ from $E_{\rm pot}$, the associated time-odd terms containing $\pmb{j_n}(\rb)$ must necessarily be dropped to avoid violation of Galilean invariance~\cite{engel1975,dobaczewski1995}.   
In this way,  the effective mass reduces to the bare mass and the potential vector vanishes identically since the Skyrme potential energy $E_{\rm pot}$ no longer depends on $\tau_n(\rb)$ and $\pmb{j_n}(\rb)$. However, the effective mass terms now contribute to the central potential
\beqy\label{eq:potential-no-effmass}
U_n(\rb) \rightarrow U_n(\rb) + \frac{\delta E_{\rm pot}}{\delta \tau_n(\rb)}\frac{\partial \tau_n(\rb)}{\partial n_n(\rb)} \, .
\eeqy 
The single-particle Hamiltonian~\eqref{eq:Hamiltonian} thus reduces to 
\beqy\label{eq:Hamiltonian-no-effmass}
h_n(\rb)= -\frac{\hbar^2}{2m_n}\pmb{\nabla}^2+U_n(\rb) + \left(\frac{\hbar^2}{2m_n^\oplus(\rb)}-\frac{\hbar^2}{2m_n}\right)[3\pi^2 n_n(\rb)]^{2/3}\, .
\eeqy
Here $m_n^\oplus(\rb)$ is the Skyrme effective mass calculated in the usual way. This effective potential is represented by the dashed line in Fig.~\ref{fig:potential}. 
With the cancellation of $\pmb{I_n}(\rb)$,  Eq.~\eqref{eq:divQv} is now satisfied exactly. In the limit of pure neutron matter at density $\rho_{n,f}$, the single-particle energies are given by 
\beqy 
\epsilon^0_{\pmb{k}}=\frac{\hbar^2 k^2}{2m_n}+U_n +\left(\frac{\hbar^2}{2m_n^\oplus}-\frac{\hbar^2}{2m_n}\right)k_F^2 \, ,
\eeqy 
therefore  
\beqy 
\pmb{\nabla}_{\pmb{k}}\epsilon^0_{\pmb{k}} = \frac{\hbar^2 \pmb{k}}{m_n} \, .
\eeqy 
It can be easily seen from Eqs.~\eqref{eq:superfluid-density-general} and \eqref{eq:average-neutron-mass-density2} that $\rho_{n,s}=\rho_{n,f}$ in this case. Unlike the original Hamiltonian~\eqref{eq:Hamiltonian} keeping $m_n^\oplus(\rb)$ but droppping $\pmb{I_n}(\rb)$, the transformation~\eqref{eq:Hamiltonian-no-effmass} thus allows to recover the correct limit in uniform neutron matter.

The neutron band structure has been recalculated for the Hamiltonian~\eqref{eq:Hamiltonian-no-effmass}.  Note that the neutron chemical potential as calculated 
using Eq.~(B21) of Ref.~\cite{pearson2012} remains unchanged with this transformation of the Hamiltonian. 
The superfluid density has been evaluated 
from Eq.~\eqref{eq:super-density-weak2} using 
the Gilat-Raubenheimer method.  For this calculation, a grid of $25\times 25\times 25$ points with 1360 cubic cells in the IBZ (65280 in the first BZ) have been considered. 
The superfluid fraction $\rho_{n,s}/\rho_{n,f}$ is found to \emph{decrease} from 7.91\% to 7.59\%.  As expected, the 
change of the superfluid fraction $\delta \rho_{n,s}/\rho_{n,f}\approx 4\%$ is small  and comparable to that in uniform neutron matter at the density $\rho_{n,f}$. 

Calculations of the superfluid fraction including $\pmb{I_n}(\rb)$ explicitly are left for future studies.

\section{Conclusion}
\label{sec:conclusion}

The neutron superfluid density in the inner crust of a neutron star has been estimated in the limit of small neutron superfluid velocities $\bar V_{n,s}\ll V_{n,L}$ from 
fully three-dimensional neutron band-structure calculations in a body-centered cubic lattice with pairing treated in the BCS approximation. 
A rigorous derivation of the superfluid density has been provided from the self-consistent  time-dependent HFB equations, thus clarifying the various assumptions made 
in Ref.~\cite{CCH05b}. For comparison,  numerical 
calculations have been carried out over a range of pairing gaps including the limit of vanishing pairing. Extensive numerical 
tests have been performed to assess the numerical precision of the code. In particular, the lattice Green function and the density of states for 
the tight-binding model have been computed, and numerical results have been compared with known analytical expressions. It has been also shown that the spatial resolution 
should be high enough to properly resolve the detailed structure of the Bloch wavefunctions inside the clusters and correctly describe the 
scattering of free neutrons by the individual nucleons in the clusters. A grid spacing $\delta x$ larger than about 1.3~fm along each Cartesian axis (about 2.3 fm along each primitive axis) leads to a spurious 
enhancement of the superfluid density, and the errors grow as a power law with $\delta x$. Setting $\delta x\approx 2.2$ fm overestimates the 
superfluid fraction by about a factor 4. 

With the new code, the neutron superfluid density is found to represent about 8\% of the density of free neutrons in the intermediate region 
of the inner crust at the average baryon density 0.03~fm$^{-3}$. This result is essentially independent of the value of the pairing gap within 
the numerical precision of the code. In the limit of homogeneous neutron-proton superfluid mixtures, as in the region beneath the crust of a 
neutron star, the HFB equations can be solved exactly. In this case, the superfluid densities were shown to be independent of the pairing gaps, 
no matter how strong the pairing, not only in the limit of small velocities but for any velocity not exceeding Landau's velocity~\cite{allard2021}.  
The present calculations suggest that this conclusion also holds for an inhomogeneous superfluid.

The suppression of the superfluid fraction challenges the classical interpretation of pulsar frequency glitches in terms of neutron 
superfluidity in neutron-star crust~ \cite{andersson2012,chamel2013,delsate2016},  and brings support to models involving an additional superfluid 
reservoir in the core.  Observations of the glitch rise in the Vela and Crab pulsars point towards the interactions of the neutron superfluid vortices 
with proton flux tubes in the outer core of neutron stars~\cite{sourie2020}.

However, the validity of the BCS approximation adopted in this work has been questioned~\cite{watanabe2017,minami2022,almirante2024}. In particular, 
Minami and Watanabe~\cite{minami2022} compared the superfluid density calculated from the HFB equation and the BCS approximation using a three-band 
one-dimensional toy model with a periodic potential given by $U(x)=2V\cos(K x)$, where $K=2 \pi/a$ and $a$ is the lattice spacing. In the region of interest, 
$\epsilon_{\rm F}\approx 18$~MeV and $\Delta\approx 1.6$~MeV therefore $\Delta/\epsilon_{\rm F}\approx 0.09$. As can be seen in 
Fig.~\ref{fig:potential}, the relative depth of the potential is about 32~MeV corresponding to $V\approx 16$~MeV and $V/\Delta \approx 10$. Looking 
at the blue curve in Fig.~2 of Ref.~\cite{minami2022}, which is the closest to the situation considered here, no significant difference would be 
expected between the BCS approximation and the full HFB equation.  
Still, full three-dimensional HFB calculations with realistic self-consistent mean fields in a body-centered cubic crystal lattice remain  
needed to draw more definite conclusions about the role of pairing on the superfluid density in this crucial region of neutron-star crusts.

Besides, the linear response theory adopted here and in previous studies certainly breaks down for superfluid velocities $\bar V_{n,s}\gtrsim V_{n,L}$. Superfluidity is not expected to disappear until $\bar V_{n,s}$ 
reaches a second critical velocity $V_{n,c} > V_{n,L}$~\cite{allard2023}. For $V_{n,L}<\bar V_{n,s}<V_{n,c}$, the superfluid becomes gapless and the superfluid 
density is further suppressed due to Cooper pair breaking~\cite{allard2023b}. Observations of transiently accreting neutron stars provide 
evidence for the existence of such gapless superfluid phase~\cite{allard2024}. 
However, the role of the crustal lattice on those characteristic velocities and the stability of the flow remains to be investigated. This calls for further 
studies of the neutron superfluid dynamics in neutron-star crusts.

\begin{acknowledgments}
This work was financially supported by FNRS (Belgium) under Grant No. PDR T.004320. 
It was performed in part at the Aspen Center for Physics, which is supported 
by National Science Foundation grant PHY-1066293. 
\end{acknowledgments}

\appendix 

\section{Reliability and accuracy of numerical integration methods}
\label{app}

\subsection{Brillouin-zone integrations}

\begin{figure}
\includegraphics[scale=0.6]{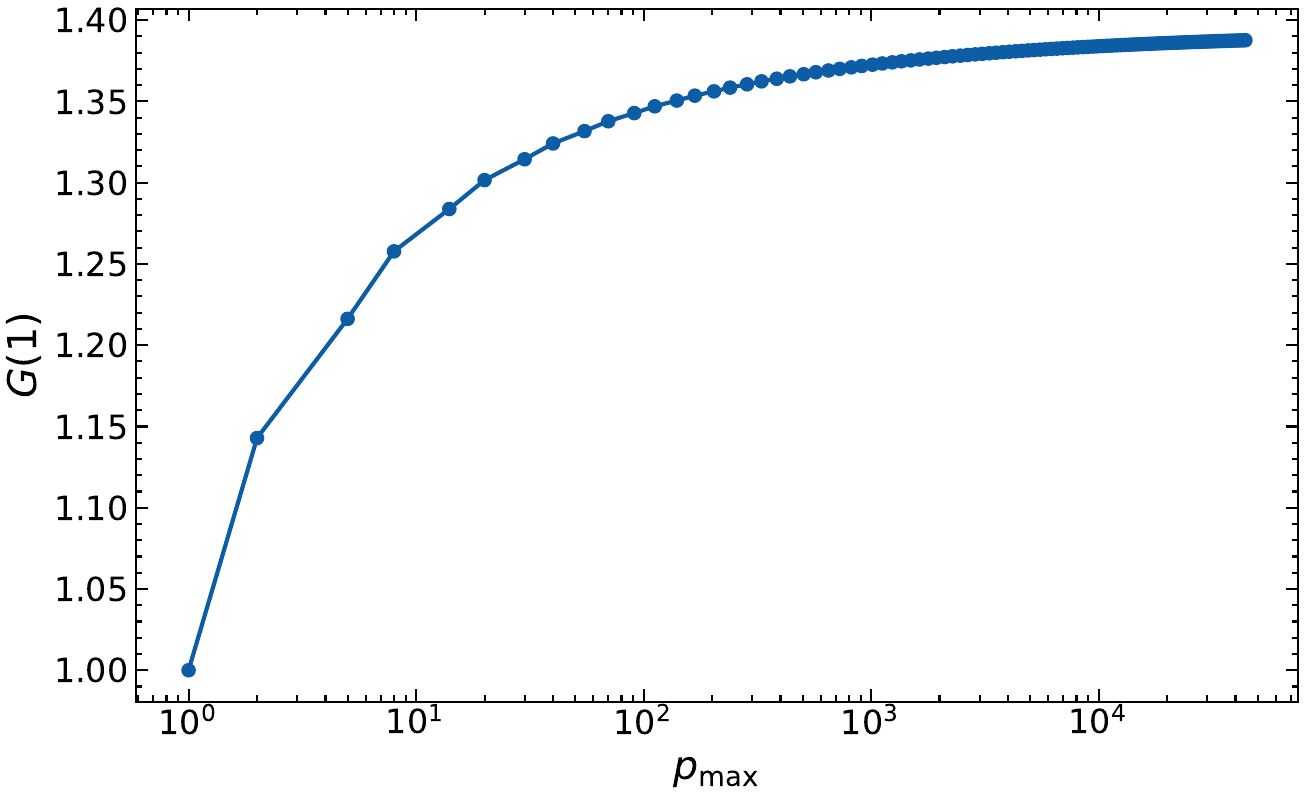}
\vskip -0.5cm
\caption{Watson's integral for a body-centered cubic lattice as evaluated using the special-point method. See text for details. }
\label{fig:Watson-integral-bcc}
\end{figure}

\begin{figure}
\includegraphics[scale=0.6]{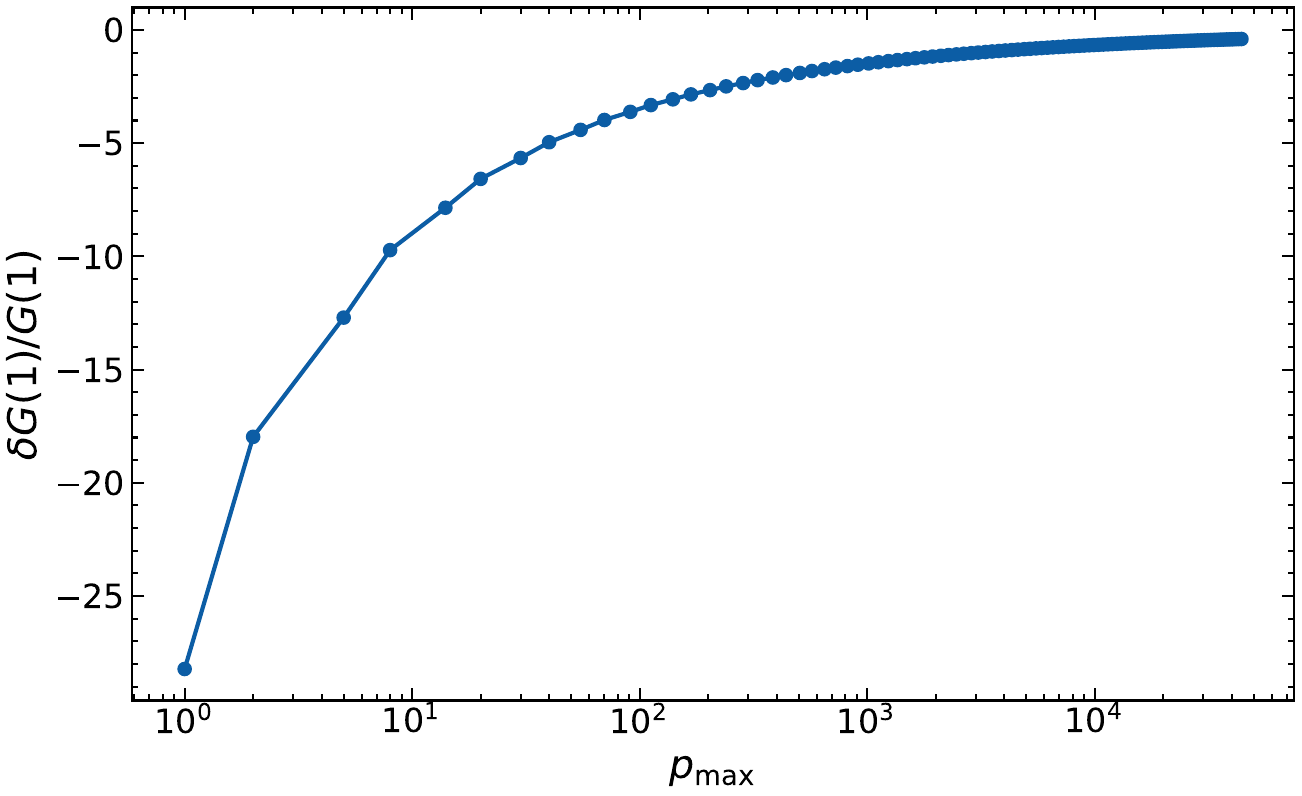}
\vskip -0.5cm
\caption{Relative deviation between the exact value of the Watson's integral for a body-centered cubic lattice and its estimate from the special-point method. See text for details. }
\label{fig:Watson-integral-bcc-err}
\end{figure}

\begin{figure}
\includegraphics[scale=0.6]{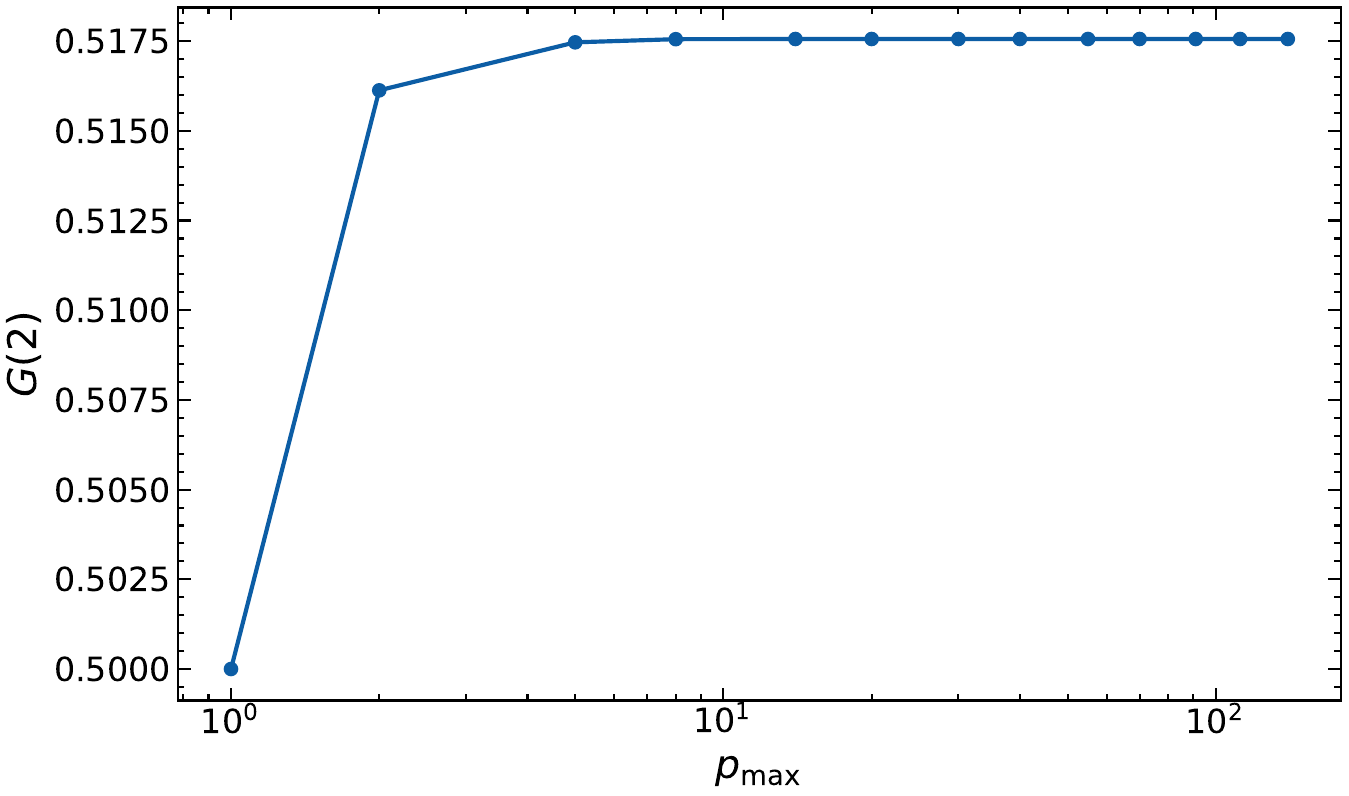}
\vskip -0.5cm
\caption{Body-centered cubic lattice Green function $G(2)$ as evaluated using the special-point method. See text for details. }
\label{fig:bcc-lattice-Green}
\end{figure}

\begin{figure}
\includegraphics[scale=0.6]{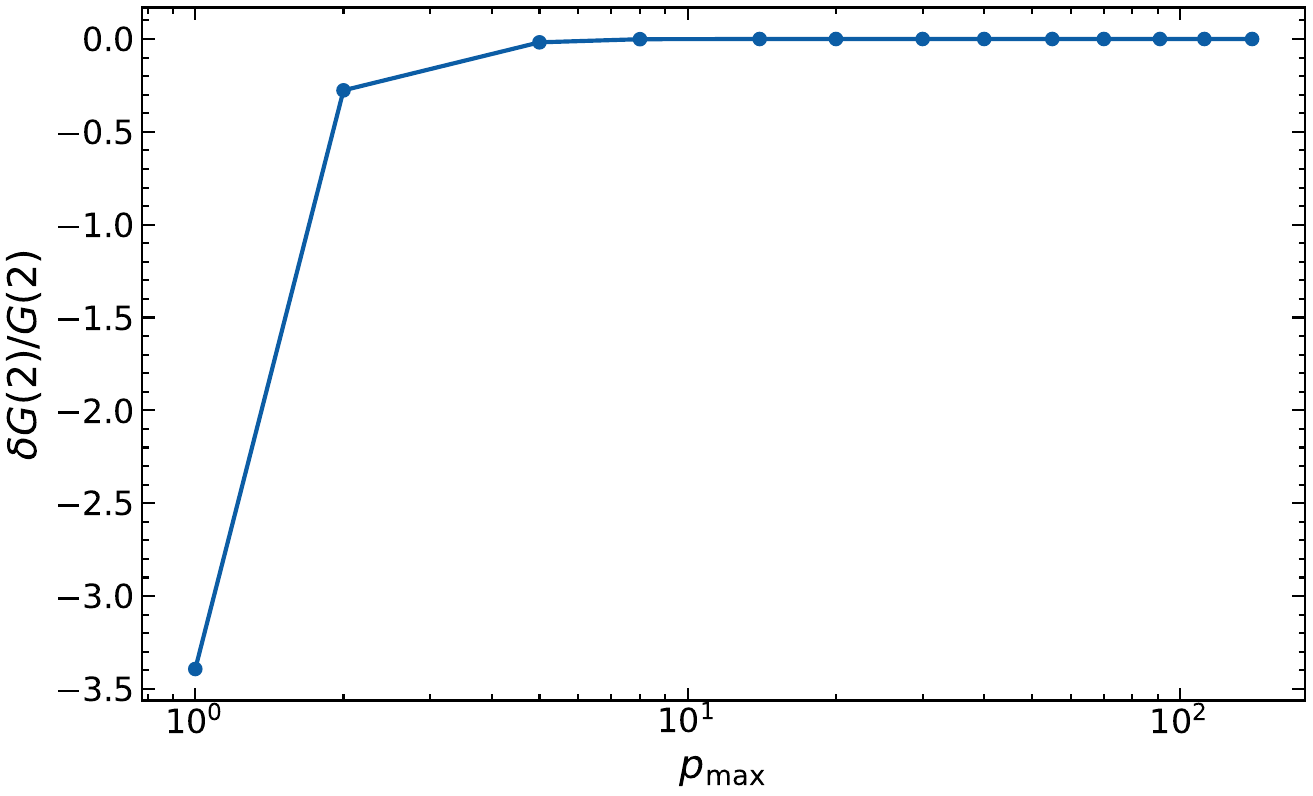}
\vskip -0.5cm
\caption{Relative deviation between the exact value of the body-centered cubic lattice Green function $G(2)$ and its estimate from the special-point method. See text for details. }
\label{fig:bcc-lattice-Green-err}
\end{figure}

To check our implementation of the special-point method and its accuracy,  
the body-centered cubic lattice Green's function has been calculated for the tight-binding model defined by~\cite{economou2006} 
\beqy\label{eq:lattice-green-function}
G(E)=\frac{1}{\Omega_{\rm BZ}}\int_{\rm BZ} \frac{d^3\pmb{k}}{E-\epsilon_{\pmb{k}} }\, , 
\eeqy  
\beqy \label{eq:tight-binding-energy}
\epsilon_{\pmb{k}}=\epsilon_0 - \epsilon_1 \cos(k_x a/2)\cos(k_y a/2)\cos(k_z a/2)\, , 
\eeqy 
where $\epsilon_0$ and $\epsilon_1$ can be expressed in terms of the tight-binding Hamiltonian. The imaginary part inside the band is 
related to the density of single-particle states. 
Without any loss of generality, one can set $\epsilon_0=0$ and $\epsilon_1=1$ (the general case can be treated by the suitable redefinitions: $E\rightarrow (E-\epsilon_0)/\epsilon_1$ and $G\rightarrow \epsilon_1 G$). From the translational symmetry of the energy bands (in reciprocal space), the integral can be equivalently carried out over the conventional (cubic) cell, whose size is $4\pi/a$ and volume is $4 \Omega_{\rm BZ}$. Recalling that the volume of the first Brillouin zone is given by $\Omega_{\rm BZ}=2 (2\pi/a)^3$, the lattice Green's function can thus be alternatively expressed as
\beqy 
G(E)=\frac{1}{4\Omega_{\rm BZ}}\int_{-2\pi/a}^{2\pi/a}dk_x\int_{-2\pi/a}^{2\pi/a}dk_y\int_{-2\pi/a}^{2\pi/a}dk_z \frac{1}{E- \cos(k_x a/2)\cos(k_y a/2)\cos(k_z a/2)}\, . 
\eeqy  
Changing variables $x\equiv k_x a/2$, $y\equiv k_y a/2$ and $z\equiv k_z a/2$, and making use of the fact that the cosine functions are even, the integral reduces to the more familiar form
\beqy 
\label{eq:lattice-green-function2}
G(E)=\frac{1}{\pi^3}\int_{0}^{\pi}dx\int_{0}^{\pi}dy\int_{0}^{\pi}dz \frac{1}{E- \cos(x)\cos(y)\cos(z)}\, . 
\eeqy  
This integral is known analytically and is given outside the band ($\vert E\vert \geq 1$) by~\cite{maradudin1960}
\beqy 
G(E)=\frac{4}{\pi^2 E}K(u)^2\, ,
\eeqy 
where 
\beqy 
u=\sqrt{\frac{1}{2}-\frac{1}{2}\sqrt{1-\frac{1}{E^2}}}\, ,
\eeqy
and $K(u)$  is the complete elliptic integral of the first kind defined by 
\beqy 
K(u)=\int_0^{\pi/2} \frac{dt}{\sqrt{1-u^2 \sin^2 t}} \, .
\eeqy 
The particular case $E=1$ is the Watson's integral~\cite{watson1939} ($\Gamma$ denotes the Gamma function): 
\beqy 
G(1)=\frac{4}{\pi^2}K\left(\frac{1}{\sqrt{2}}\right)^2=\frac{1}{4\pi^3}\Gamma\left(\frac{1}{4}\right)^4\approx 1.3932\, . %039297\, .
\eeqy  
Equation~\eqref{eq:lattice-green-function} (with $\epsilon_0=0$ and $\epsilon_1=1$) has been evaluated numerically using the special-point method. Since $G(E)$ is independent of $a$, as shown in Eq.~\eqref{eq:lattice-green-function2}, one can set $a=1$. Note that the integrand in the lattice Green function $G(1)$ may be singular. The Watson's integral thus provides a more stringent test of the integration scheme than for the lattice Green function for any other value $\vert E\vert > 1$. Numerical estimates obtained by increasing the number of special points are displayed in Fig.~\ref{fig:Watson-integral-bcc}. Using a single point~\cite{baldereschi1973} leads to a value fairly close to the exact result with a relative deviation of about 28\% only. As shown in Fig.~\ref{fig:Watson-integral-bcc-err},  the precision improves significantly by including a few more points: the error thus drops below 10\% with 8 points. Further increasing the number of points leads to a much slower convergence towards the exact result: an estimate with a 1\% precision thus requires about 3000 points. This can be traced back to the singularity of the integrand. Indeed, evaluation of the body-centered cubic lattice Green function for $E=2$ leads to an extremely fast convergence, as can be seen in Figs.~\ref{fig:bcc-lattice-Green} and \ref{fig:bcc-lattice-Green-err}: a precision of 3\% is reached with a single point and the error falls below 0.3\% by adding just another point, the exact result to 5 significant digits being $G(2)\approx 0.57915$.

%--------------------------------------------------------------------------------------------------------------------------------
%---------------------------------------------------------------------------------------------------------------------------------

%--------------------------------------------------------------------------------------------------------------------------------
%---------------------------------------------------------------------------------------------------------------------------------

%--------------------------------------------------------------------------------------------------------------------------------
%---------------------------------------------------------------------------------------------------------------------------------

%--------------------------------------------------------------------------------------------------------------------------------
%---------------------------------------------------------------------------------------------------------------------------------

\subsection{Fermi surface integrations}

\begin{figure}
	\includegraphics[scale=0.6]{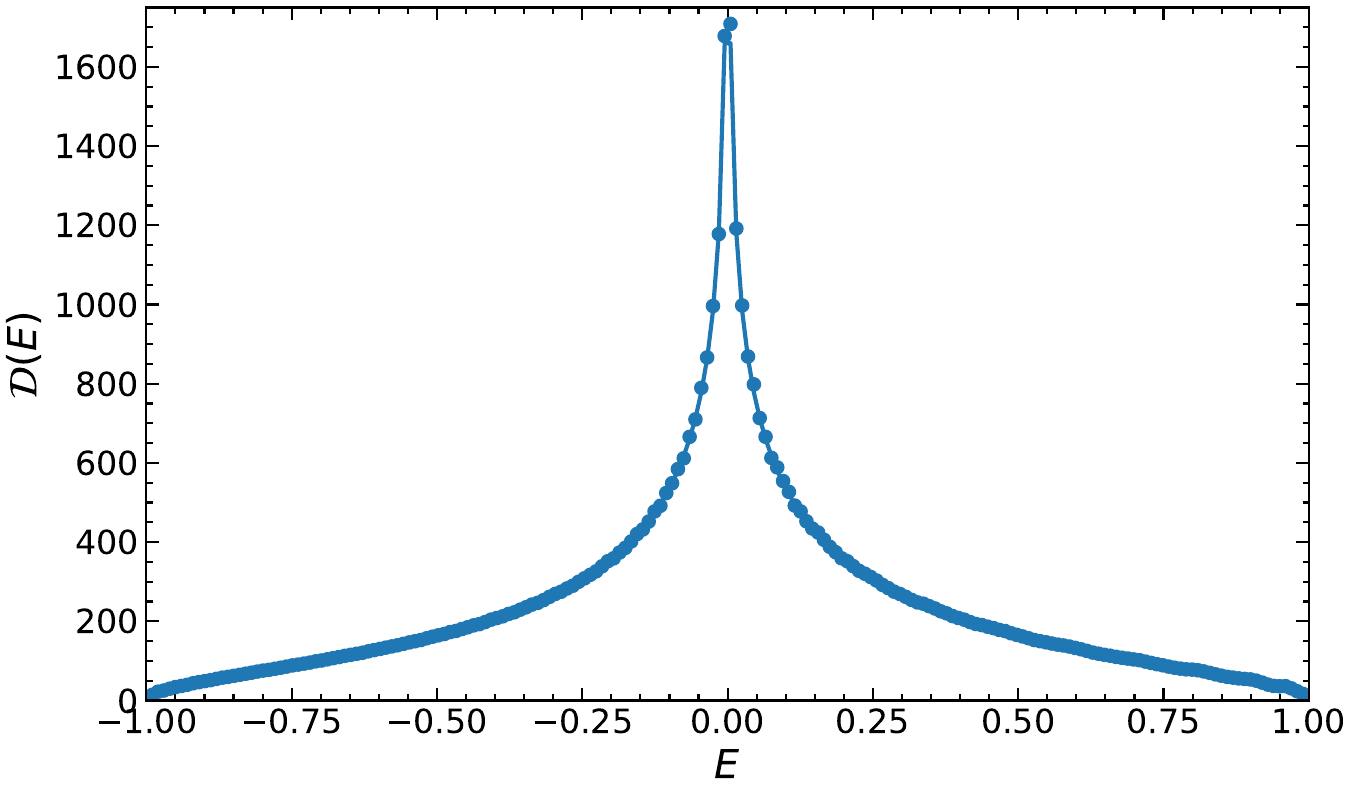}
	\vskip -0.5cm
	\caption{Density of states for the tight-binding model as calculated using the Gilat-Raubenheimer method with 1360 cubic cells (dots) and the fit (solid line) of Ref.~\cite{jelitto1969}. }
	\label{fig:dos-bcc-1360}
\end{figure}

To test our implementation of this method, the density of states has been calculated for the tight-binding model defined by 
\beqy 
\mathcal{D}(E)=\int_{\rm BZ} d^3\pmb{k}\,  \delta( E - \epsilon_{\pmb{k}})=\int \frac{ d\mathcal{S}}{\vert\pmb{\nabla}_{\pmb{k}}\epsilon_{ \pmb{k}}\vert } \approx \sum_{c=1}^{c_{\rm max}}  \frac{w_c\, \mathcal{S}_c}{\vert\pmb{\nabla}_{\pmb{k_c}}\epsilon_{ \pmb{k_c}}\vert }
\eeqy 
with the band energy given by Eq.~\eqref{eq:tight-binding-energy}. As shown in Fig.~\ref{fig:dos-bcc-1360}, results using 1360 cubic cells in the irreducible domain (65 280 cells in the first Brillouin zone) are found to be in excellent agreement with the interpolating formula (39c) of Ref.~\cite{jelitto1969} over the whole spectrum, including in particular the Van Hove singularity at $E=0$ corresponsing to critical points for which  $\vert\pmb{\nabla}_{\pmb{k}}\epsilon_{ \pmb{k}}\vert=0$.

%--------------------------------------------------------------------------------------------------------------------------------
%---------------------------------------------------------------------------------------------------------------------------------

\bibliography{references.bib}
\end{document}